\documentclass[twocolumn,preprintnumbers,amsmath,amssymb,superscriptaddress,noshowpacs,showkeys]{revtex4-1}

\usepackage{graphicx}
\usepackage{dcolumn}
\usepackage{bm}
\usepackage[T2A]{fontenc}
\usepackage[utf8]{inputenc}
\usepackage[english]{babel}
\usepackage{multirow}
\usepackage{booktabs}
\usepackage{textcomp}
\usepackage{xcolor}
\usepackage{soul}
\usepackage{gensymb}
\usepackage{natbib}
\usepackage{float}
\usepackage{caption}
\usepackage{subcaption}
\usepackage[normalem]{ulem}

\usepackage{hyperref}
\hypersetup{
	pdftitle={Acceleration Potential and Density Profile of Secondary Plasma in the Magnetosphere of Orthogonal Pulsars},
	pdfauthor={Istomin A., Kniazev F., Beskin V.}
}

\allowdisplaybreaks

\begin{document}

\title{Acceleration Potential and Density Profile of Secondary Plasma in the Magnetosphere of Orthogonal Pulsars}

\author{\firstname{A.Yu.}~\surname{Istomin}}
\email{istomin.aiu@phystech.edu}
\affiliation{Moscow Institute of Physics and Technology (National Research University), Dolgoprudny, Moscow oblast, 141701 Russia}
\affiliation{Lebedev Physical Institute, Russian Academy of Sciences, Moscow, 119333 Russia}

\author{\firstname{F.A.}~\surname{Kniazev}}
\affiliation{Moscow Institute of Physics and Technology (National Research University), Dolgoprudny, Moscow oblast, 141701 Russia}
\affiliation{Lebedev Physical Institute, Russian Academy of Sciences, Moscow, 119333 Russia}

\author{\firstname{V.S.}~\surname{Beskin}}
\affiliation{Lebedev Physical Institute, Russian Academy of Sciences, Moscow, 119333 Russia}
\affiliation{Moscow Institute of Physics and Technology (National Research University), Dolgoprudny, Moscow oblast, 141701 Russia}

\date{\today}
\begin{abstract}
A new method for determining the accelerating potential above the polar caps of radio pulsars with an arbitrary inclination angle of the magnetic axis to the rotation axis has been proposed. The approach has been based on the concept of a vacuum gap, the height and shape of the upper boundary of which are found self-consistently together with the solution of the corresponding Poisson equation. In turn, information about the accelerating potential has made it possible to determine the transverse profiles of the secondary plasma density. It has also been shown that the effect of inverse Compton scattering on the considered processes is insignificant.
\end{abstract}

\keywords{radio pulsars, accelerating potential, secondary plasma}

\maketitle
\onecolumngrid
This is an author's translation of the original paper: Astronomicheskii Zhurnal, 2025, V. 102, N. 1, pp. 53-65, \\
DOI: 10.31857/S0004629925010056
\vspace{8pt}
\twocolumngrid

\section{Introduction}
The data accumulated in recent years by the MeerKAT \cite{MeerKAT_XI, MeerKAT_VI} and FAST \cite{FAST682} radio telescopes made it possible to significantly refine the characteristics of the average intensity and polarization profiles of pulsar radio emission. This, in turn, required the creation of a more detailed theory of the generation and propagation of radiation in the magnetospheres of neutron stars. For instance, the “hollow cone” model \cite{radhakrishnan69, O&S} which is most often used to explain the properties of the observed radiation profiles, turns out to be overly simplified since it does not take into account propagation effects in the neutron star magnetosphere, such as refraction, cyclotron absorption, and the effect of limiting  polarization \cite{beskinphilippov2012}. 

In order to study all of the above effects, it is necessary to have a quantitative model of the density distribution of the secondary electron-positron plasma flowing along the open field lines of the radio pulsar. At the same time, the main mechanism of plasma generation in the polar regions of a neutron star is single-photon conversion of $\gamma$-quanta emitted by primary electrons and/or positrons accelerated in the region of a non-zero longitudinal electric field. That is why, in order to create such a model, it is necessary to know the three-dimensional structure of the accelerating potential in the region above the polar cap of the neutron star, which is the main goal of this paper.

Although the classical Ruderman--Sutherland vacuum gap model \cite{RS} serves as a good starting point for solving this problem, in its original formulation, it contains a significant uncertainty: the mean free path
of photons and, therefore, the geometry of the vacuum gap itself depends on the magnitude of the accelerating potential. Therefore, this model can only be used for sufficiently fast radio pulsars, when the gap height is much smaller than the radius of the polar cap, and the corresponding electric field can be considered to be uniform. Moreover, this model cannot be used for orthogonal pulsars, the inclination angle of the magnetic axis to the rotation one of which is close to 90\textdegree. Indeed, as was shown in \cite{UFN}, the accelerating potential of such pulsars does not have axial symmetry and, accordingly, cannot be described within the framework of the classical model. As for the numerous studies performed over the last ten years within the framework of particles in cells (PIC) \cite{Tim2010, TimArons2013, philippovspitkovsky2014, TchPhS16, TimHar2015}, the overwhelming majority of them (with the exception, perhaps, of \cite{Cruz}) were not aimed at studying the spatial distribution of secondary plasma above the polar cap.

In this paper, we propose a way to formulate and solve the problem of determining the spatial structure of the accelerating potential correctly. Our method does not assume that gap height $H_{gap}$ is small, compared to the $R_0$ radius of the polar cap, and that the potential possesses axial symmetry, which makes it applicable to orthogonal pulsars. The last remark is especially important, since the possibility of observing both magnetic poles makes these objects particularly informative \cite{ABP2017}. Besides, due to the small magnitude and sign-variability of the Goldreich--Julian charge density within the polar cap, such pulsars are particularly sensitive to the choice of the plasma generation model and the strength of the magnetic field.

In Section \ref{sec:basic_eq}, the mathematical statement of the considered problem is formulated. Section \ref{sec:acc_potential} is devoted to the calculation of the vacuum gap height,the solution of the Poisson equation using Physics Informed Neural Networks, and discussion of the results of these calculations for both non-orthogonal and orthogonal pulsars. Section \ref{sec:Dencity} is devoted to the discussion of the secondary plasma generation model and the results of its application to previously determined potentials. In Section \ref{sec:Conclusion}, the results of this study are summarized. 

\section{Basic equations}
\label{sec:basic_eq}
Although the results of numerical simulations indicate that the particle generation is essentially a time-dependent process \cite{TimHar2015, PhSC15}, we chose the stationary vacuum gap model \cite{RS} as a starting point. Indeed, since, according to \cite{TimHar2015, PhSC15}, the plasma periodically completely leaves the magnetosphere, the Ruderman--Sutherland model is suitable for description of the initial stage of secondary plasma generation. However, in this paper, as was previously said, we are not going to assume the condition $H_{gap} \ll R_0$, where
\begin{equation}
    R_0 = f^{1/2}_* \sqrt{\frac{\Omega R}{c}} R
    \label{eq:polaR_cap}
\end{equation}
is the polar cap radius. From here on $R$, $\Omega$ and $\chi$ are the star radius, the angular velocity of its rotation, and the inclination angle of the magnetic axis to the rotation one. The quantity $f_* \approx 1.59 (1 + \sin^2{\chi})$ is a dimensionless area of the polar cap \cite{bgi83, Gralla}.

We now move on to the calculation of the electric potential $\psi$ in the vacuum region above the polar cap. In a steady state, the Poisson equation for a rotating neutron star has the form \cite{RS}:
\begin{equation}
    \Delta \psi = 4 \pi (\rho_{\mathrm{e}} - \rho_{\mathrm{GJ}}),
    \label{eq:PoissonGeneral}
\end{equation}
where $\rho_{\mathrm{e}}$ --- charge density in the magnetosphere and
\begin{equation}
    \rho_{\mathrm{GJ}} = - \frac{\mathbf{\Omega}\mathbf{B}}{2 \pi c}  
    \label{eq:rho_GJ}
\end{equation}
-- is the Goldreich-Julian charge density \citep{GJ}, necessary for screening the longitudinal electric field. 

The problem can be simplified if one notes that on the scale of the gap height, the region of open field lines differs little from the cylindrical one. This makes it possible to neglect the curvature of the field lines when choosing the computation domain. Thus, we can introduce cylindrical coordinates $r_\mathrm{m}$ , $\phi_\mathrm{m}$ and $z$ with the center on the dipole axis, where normalization of quantities $r_\mathrm{m}$ and $z$ to the polar cap radius $R_0$ \eqref{eq:polaR_cap} is chosen. As an example, the surface separating the regions of open and closed field lines is determined by the condition: $r_m = 1$. However, the curvature of the field lines must be taken into account in the right-hand side of equation \eqref{eq:PoissonGeneral}, since at angles $\chi$ close to 90\textdegree, the axisymmetric (and independent of the curvature of the magnetic field lines) term tends to zero. 

In this case, the equation for accelerating potential $\psi(r_{\mathrm{m}}, \varphi_{\mathrm{m}}, z)$ in a vacuum region ($\rho_e \ll \rho_{\mathrm{GJ}}$) can be written down in the following way:
\begin{eqnarray}
\frac{1}{r_{\mathrm{m}}} \frac{\partial}{\partial r_{\mathrm{m}}} 
\left(r_{\mathrm{m}}\frac{\partial\psi}{\partial r_{\mathrm{m}}}\right) 
+ \frac{1}{r_{\mathrm{m}}^2} \frac{\partial^2 \psi}{\partial \varphi_{\mathrm{m}}^2} 
+ \frac{\partial^2 \psi}{\partial z^2}
\nonumber 
=\\ - 2 \, \frac{\Omega B_{0} R_0^2}{c} 
\left(\cos\chi + \frac{3}{2} \, \frac{R_{0}}{R} \, r_{\mathrm{m}}\sin\chi \sin\varphi_{\mathrm{m}}\right).
\label{eq:Poisson}
\end{eqnarray}
Here we have used a well-known expression for the cosine of the angle between the rotation axis and dipolar magnetic field 
\begin{equation}
    \cos \theta_{\rm b} \approx \cos \chi + (3/2) (R_{0}/R) \,  r_{\mathrm{m}}\sin\chi \sin\varphi_{\mathrm{m}},
\end{equation}
as throughout this paper, magnetic field geometry is assumed to be dipole, although the proposed method can be generalized to other configurations. If one assumes that the gap height $H_{\mathrm{gap}} = H_{\mathrm{gap}} (r_{\mathrm{m}}, \phi_{\mathrm{m}})$ is a known function, the border conditions for the equation \eqref{eq:Poisson} can be written down as follows:
\begin{eqnarray}
   \psi(r_{\mathrm{m}}, \varphi_{\mathrm{m}}, z = 0) = 0,
\label{eq:bc0} \\
\psi(r_{\mathrm{m}} = 1, \varphi_{\mathrm{m}}, z) = 0,
\label{eq:bc1} \\
\frac{\partial \psi}{\partial z}(r_{\mathrm{m}}, \varphi_{\mathrm{m}}, z)\bigg |_{z = H_{\mathrm{gap}} (r_{\mathrm{m}}, \varphi_{\mathrm{m}})}= 0.
\label{eq:bc2}
\end{eqnarray}
In other words, we set zero border conditions on the polar cap surface and on the separatrix, and we also require the electric field on the upper border of the gap $z = H_{\mathrm{gap}}$ to be zero \citep{TimHar2015, Novoselov}.

However, in order to know $H_{\mathrm{gap}}$ function one should consider the pair plasma generation processes, which is impossible without information about primary particles energies. These energies in turn depend on $H_{\mathrm{gap}}$ function, as accelerating potential $\psi$ depends on the gap height due to border conditions $\psi = \psi[H_{\mathrm{gap}}]$. Therefore, the correct treatment of the problem is simultaneous and self-consistent computation of: $\psi (r_{\mathrm{m}}, \phi_{\mathrm{m}}, z)$ and $H_{\mathrm{gap}}(r_{\mathrm{m}}, \phi_{\mathrm{m}})$, which satisfy the condition
\begin{equation}
    \frac{\partial \psi(r_{\mathrm{m}}, \phi_{\mathrm{m}}, z)}{\partial z} \bigg |_{z = H_{\mathrm{gap}}[\psi]} = 0.
\end{equation}
In order to solve this problem we developed an iterative scheme: knowing $H_{\mathrm{gap}}^{(i)}$ we obtain $\psi^{(i)}$ which satisfies \eqref{eq:bc2}, and after that recalculate the gap height with a fixed potential. Then this new gap height is used on the next step of the procedure. We repeat these steps until we converge to certain solutions for $\psi$ and $H_{gap}$:
\begin{align}
    H_{\mathrm{gap}}^{(i)} &\rightarrow \psi^{(i)}, \nonumber \\
    H_{\mathrm{gap}}^{(i + 1)} &= w H_{\mathrm{gap}}[\psi^{(i)}] + (1 - w)  H_{\mathrm{gap}}^{(i)}. 
    \label{eq:iterations}.
\end{align}
Here $0 < w < 1$ --- some weighting factor for better conversion.
However, the exact implementation of this scheme is rather tricky. On each step of the iterative scheme \eqref{eq:iterations} one should not only determine the gap height $H_{\mathrm{gap}}[\psi]$, but solve the Poisson equation \eqref{eq:Poisson} with border conditions \eqref{eq:bc0}, \eqref{eq:bc1}, \eqref{eq:bc2}, the last of which is set on a surface with a shape changing from one iteration to another. Therefore, we should discuss the implementation of our method in much more detail, which is done in the next section. 
\section{Accelerating potential}
\label{sec:acc_potential}
\subsection{Curvature radiation}

In order to determine the gap height, it is necessary to consider the absorption of primary photons in an ultra-strong magnetic field and the mechanism of their generation. Usually, the source of gamma-quanta of the required energies is considered to be curvature radiation of primary particles. In this case, for the gap height $H_{\mathrm{gap}}$, we can write down
\begin{equation}
    H_{\mathrm{gap}}(r_{\mathrm{m}}, \phi_{\mathrm{m}}) = \min_{l_{\mathrm{e}}}(l_{\mathrm{e}} + l_{\gamma} (l_{\mathrm{e}})),
    \label{eq:minimisation}
\end{equation}
where $l_{\mathrm{e}}$ is the distance traveled by the primary particle before emitting a photon. For the curvature radiation, as the primary particle energy increases, both the number of emitted photons and their characteristic energy increase too. Therefore, we can assume that a primary photon can be emitted at any point of the particle trajectory and impose no bounds on $l_{\mathrm{e}}$. 

Next, for dimensionless free path length for a photon with an energy $\epsilon_{\gamma}$, that moves in the magnetic field
of a neutron star $B$, we have the following expression \citep{Sturrock, RS}:
\begin{equation}
    l_{\gamma} \approx \frac{8}{3}\frac{1}{\Lambda} \frac{B_{\mathrm{cr}}}{B}\frac{R_{\mathrm{c}}(r_{\mathrm{m}}, z)}{\epsilon_{\gamma} (l_{\mathrm{e}})} .
    \label{eq:l_gamma}
\end{equation}
From here on, all energy quantities are normalized to $m_{\mathrm{e}} c^2$,
$B_{\mathrm{cr}} = m_{\mathrm{e}}^2 c^3 / \hbar e \approx 4.4 \cdot 10^{13} \;\text{Gs}$ --- Schwinger magnetic field, $R_{\mathrm{c}}(r_{\mathrm{m}} , z) \approx (4/3)R^2/r_{\mathrm{m}}$ --- field lines curvature radius and \mbox{$\Lambda = \Lambda_0 - 3 \ln{\Lambda_0}$,} where
\begin{equation}
    \Lambda_0 = \ln \left[\frac{e^2}{\hbar c} \frac{\omega_B R_{\mathrm{c}}}{c}\left(\frac{B_{\mathrm{cr}}}{B}\right)^2 \epsilon_{\gamma}^{-2}\right] \sim 20
    \label{eq:Lambda}
\end{equation}
is the dimensionless parameter which weakly (logarithmically) depends on pulsar parameters and photon energy \citep{PaperI}. 

Finally, in order to simplify acceleration potential calculation, we use monoenergetic approximation for curvature radiation spectrum.
\begin{equation}
    \epsilon_{\gamma} = \epsilon_{c} = \frac{3 \lambda_{\mathrm{e}} \gamma_{\mathrm{e}}^3(l_{\mathrm{e}})}{2 R_{\mathrm{c}}(r_{\mathrm{m}}, z)}.
    \label{eq:mono_spectrum}
\end{equation}
Indeed, although photon free path length \eqref{eq:l_gamma} obviously depends on its energy, in Section \ref{subsec:axisym_potential} it will be shown that the dependence of accelerating potential on numerical coefficient in expression \eqref{eq:mono_spectrum} is much weaker.  
Here $\lambda_{\mathrm{e}} = \hbar / m_{\mathrm{e}} c$ is reduced Compton wavelength, and $\gamma_{\mathrm{e}}$ is $\gamma$-factor of primary particle which is determined from the equation of motion:
\begin{equation}
   m c^2 \, \frac{d \gamma_{\mathrm{e}} }{d l} = e E_{\parallel} - F_{\mathrm{CR}}  - F_{\mathrm{IC}}.
    \label{eq:EOM}
\end{equation}
Here $E_{\parallel} = d\psi / dz$ is the parallel electric field and the last two terms correspond to the losses due to curvature radiation and inverse Compton scattering. However, contribution of $F_{\mathrm{CR}}$ is significant only at $\gamma \sim 10^8$, which makes it possible to neglect it for most pulsars. The Compton scattering losses can also be shown to be negligible if one compares the corresponding free path lengths of primary particles which was done in the subsequent subsection ~\ref{subsec:ICS}. Thus, in most cases, $\gamma$-factor of primary particle can be determined, without solving differential equation ~\eqref{eq:EOM} directly, but using its solution in the absence of losses:

\begin{equation}
    \gamma_{\mathrm{e}}(l) = \frac{e \psi(l)}{m c^2}.
\end{equation}

\subsection{Inverse Compton scattering}
\label{subsec:ICS}

In a number of papers, it was stated that for a wide range of pulsar parameters, the key mechanism which determines the gap height may be inverse Compton scattering (both non-resonant and resonant) of thermal photons on ultra-relativistic primary particles \citep{HibAronsPulsarDeath, ZhangCompton}. The key difference of this process from the curvature radiation is that the increase in primary particle energy results in the increase in Compton photons energy and \textit{decrease} in their production rate. Therefore, primary particles with an energy enough to emit a pair-producing $\gamma$-quantum can have a negligible probability of actually experiencing a scattering event. It means that we must explicitly take into account finiteness of primary particle free path length, which limits $l_{\mathrm{e}}$ in equation \eqref{eq:minimisation} from below. 

If one analyzes the dependence of this quantity on energy, the effect of inverse Compton scattering on the considered processes can be estimated. Assuming the photon emission spectrum to be thermal and isotropic in the range of $\mu_{\mathrm{min}} \leq \mu \leq 1$ ($\mu$ --- cosine of the impact angle) we can write down the following expression for a scattering event frequency \citep{Blumenthal_Gould}:
\begin{eqnarray}
    R(\gamma_{\mathrm{e}}) = \int \limits_{0}^{+\infty} d \epsilon \int \limits_{\mu_{\mathrm{min}}}^{1} d \mu \frac{c}{4 \pi^3 \lambda_{\mathrm{e}}^3} \frac{\epsilon^2}{e^{\epsilon / T} - 1} \cdot \nonumber \\ (1 - \beta \mu) \sigma_{\mathrm{tot}}(\gamma, \epsilon^{\prime} (\epsilon, \mu)).
    \label{eq:ICS_free_path}
\end{eqnarray}
Here $\epsilon$ is the photon energy before scattering in the pulsar reference frame (in $m_{\mathrm{e}} c^2$ ), $\epsilon^{\prime}$ is the photon energy in the electron reference frame and $\sigma_{\mathrm{tot}}$ --- total scattering cross section. 

As the primary particle is almost instantly accelerated up to $\gamma \gtrsim 10^5$, for the non-resonant photons relativistic Klein–Nishina cross section should be used. At the same time, since the cyclotron resonance takes place at energies \mbox{$\epsilon^{\prime} = \epsilon_{\mathrm{B}} = B / B_{\mathrm{cr}} \ll 1$}, in the non-resonant regime one can neglect external magnetic field and use the following expression: ($\sigma_{\mathrm{T}}$ --- Thomson cross section) \citep{BLP}
\begin{equation}
    \sigma_{\mathrm{NR}} = \frac{3 \sigma_{\mathrm{T}}}{8} \frac{1}{\epsilon^{\prime}} \left[\left(1-\frac{2}{\epsilon^{\prime}}-\frac{2}{\epsilon^{\prime 2}}\right) \ln (1+2 \epsilon^{\prime} )  +\right.D \left. \frac{1}{2}+\frac{4}{\epsilon^{\prime}}-\frac{1}{2(1+2 \epsilon^{\prime})^2}\right].
    \label{eq:sigma_NR}
\end{equation}

Due to the same reason, resonant cross section can be considered in the Thomson limit (taking the external magnetic field into account). Then, the cross section can be calculated as follows: \citep{Herold1979}
\begin{equation}
    \sigma=\frac{\sigma_{\mathrm{T}}}{2}\left[\frac{u^2}{(u+1)^2}+\frac{u^2}{(u-1)^2+a^2}\right].
\end{equation}
Here $u = \epsilon^{\prime} / \epsilon_{\mathrm{B}}$, $a = 2 \alpha  / 3 \epsilon_{\mathrm{B}}$ and $\epsilon_{\mathrm{B}} = B / B_{\mathrm{cr}}$. Isolating the resonant part in this expression, we can write:
\begin{equation}
    \sigma_{\mathrm{R}} = \frac{3 \pi}{4 \alpha_{\mathrm{e}}} \sigma_{\mathrm{T}} \delta(\epsilon^{\prime} - \epsilon_{\mathrm{B}})
    \label{eq:sigma_R}.
\end{equation}

For our purposes, it is sufficient to study the dependence of the primary particle free path length on its energy without considering acceleration details, which means that we can simply put $l_{\mathrm{e}} (\gamma_{\mathrm{e}}) = c / R(\gamma_{\mathrm{e}})$.
For the non-resonant scattering we numerically integrate \eqref{eq:ICS_free_path} using \eqref{eq:sigma_NR}, while for the resonant case an analytical expression can be obtained \citep{Dermer1990}:
\begin{equation}
    l_{\mathrm{e}}^R = \frac{\lambda_{\mathrm{e}} \gamma_{\mathrm{e}}^2}{\epsilon_{\mathrm{B}}^2 \Theta \alpha_{\mathrm{e}} (-\ln{( 1- e^{-w})})},
\end{equation}
where $\Theta = T / m_e c^2$, and
\begin{equation}    
    w = \frac{\epsilon_{\mathrm{B}}}{\gamma_{\mathrm{e}} \Theta (1 - \beta \mu_{\mathrm{min}})} .
\end{equation}

\begin{figure}[H]
\centering
\includegraphics[width=0.95\linewidth]{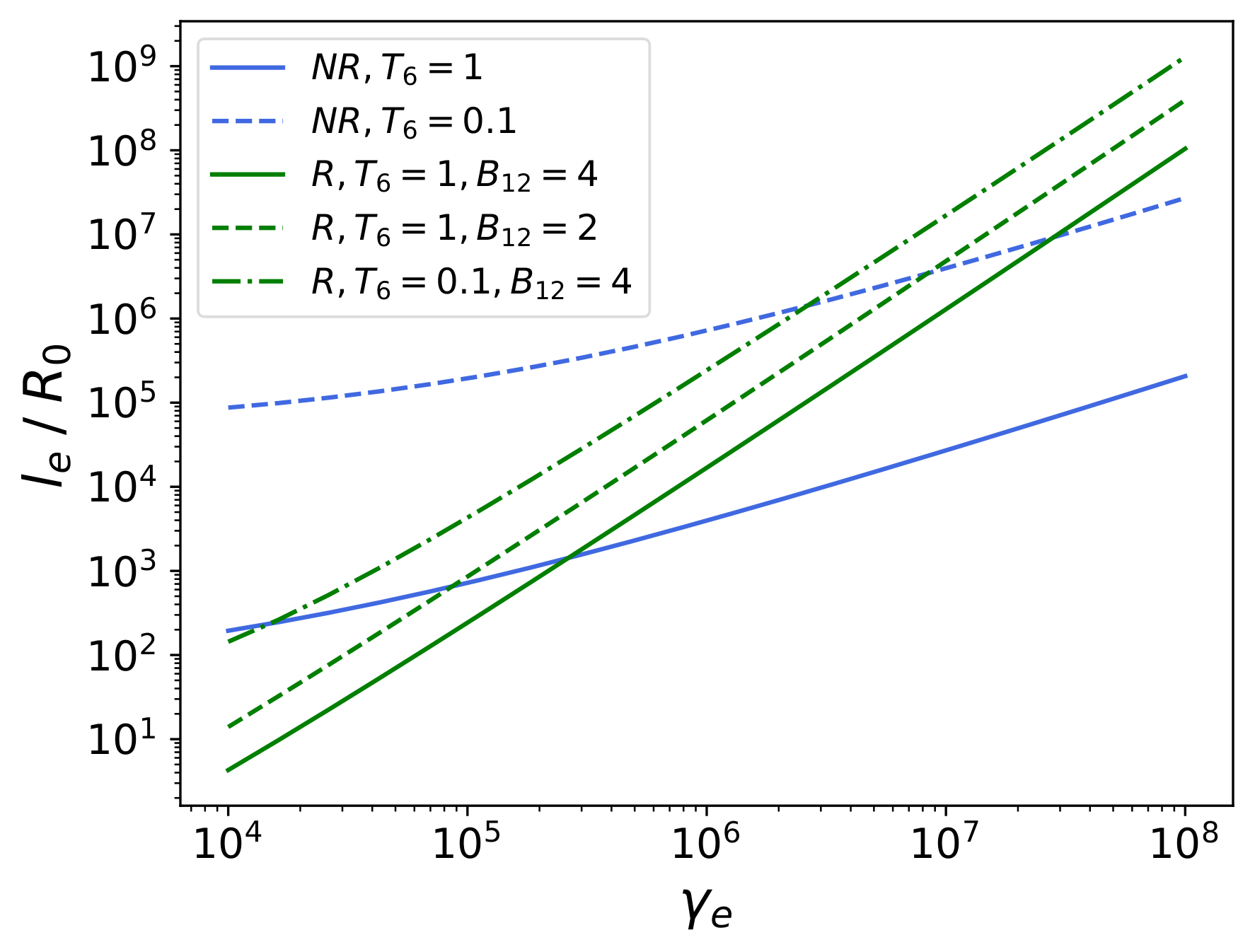}
\caption{Dependence of the mean free path of a primary particle on its $\gamma$-factor for various parameters of the pulsar. Green color (three parallel lines) corresponds to resonant scattering $(R)$, and blue color (two gentle curves) corresponds to non-resonant one $(NR)$.}
\label{fig:l_e}
\end{figure}

Then, we compare obtained free path lengths with a polar cap radius, which is a characteristic gap height scale in case of predominance of the curvature radiation mechanism. The results are presented in Fig.~\ref{fig:l_e}. As one can see, in the non-resonant regime, even at temperature $10^6$ K free path lengths are comparable to the polar cap radius only when $\gamma \lesssim 10^2$, which allows us to neglect this process throughout accelerating potential calculation. In the resonant regime, the answer is not so clear since at $\gamma \sim 10^4$,  the energy of scattered photons may be sufficient to produce a pair near the star. However, according to our calculations, this process can also be neglected when corresponding parameters are within the ranges observed for orthogonal pulsars.

\subsection{Solution of the Poisson Equation by a PINN Method}
\label{sec:PINN}
The next step towards the implementation of the iterative scheme \eqref{eq:iterations} is the solution of the Poisson equation in the vacuum region \eqref{eq:Poisson}. Although one can obtain an analytical solution for this equation in the form of an infinite series~\citep{PaperI}, this approach is not preferable in our case. The reason is that the system of linear equations, which determines expansion coefficients, has an extremely large condition number in a parameter space of interest. Traditional grid-based methods are also poorly suited for our problem due to the complex computation domain with a variable shape. 

In this regard, we decide to use an alternative approach to solve partial differential equations: Physics-Informed Neural Networks (PINN). This method is based on the usage of neural networks to obtain approximate solutions of PDEs, and is currently being increasingly applied in many areas of physics and astrophysics such as, cosmology~\citep{Chantada2023}, radiation transfer  \citep{RadiativeTransportPinn} and the theory of pulsars magnetosphere~\citep{Pons2023, Contopoulos2024}.
PINN is an ordinary neural network of some kind (usually a fully connected neural network), the input parameters of which are independent variables $\mathbf{x}$ and the output is a function $\mathcal{N} (\mathbf{x})$, which approximately satisfies the equation $\hat{L}[ \mathcal{N} (\mathbf{x})] = 0$, where $\hat{L}$ is some differential operator. In order to train the neural network in such a way, one can use the norm of PDE residuals as a loss function:  $||\hat{L}[ \mathcal{N} (\mathbf{x})]|| \rightarrow 0$. It is important that no grid is needed to calculate corresponding derivatives, as they can be obtained directly in the process of backpropagation.  Thus, during the training, the neural network will converge to one of the solutions of the corresponding differential equation. As for the boundary conditions, the simplest way to take them into account is to add additional terms to the loss function with some weight coefficients. However, within this approach, it is needed to balance the weights to achieve minimization of all terms individually.
\begin{figure}[h]
\centering
\includegraphics[width=0.9\linewidth]{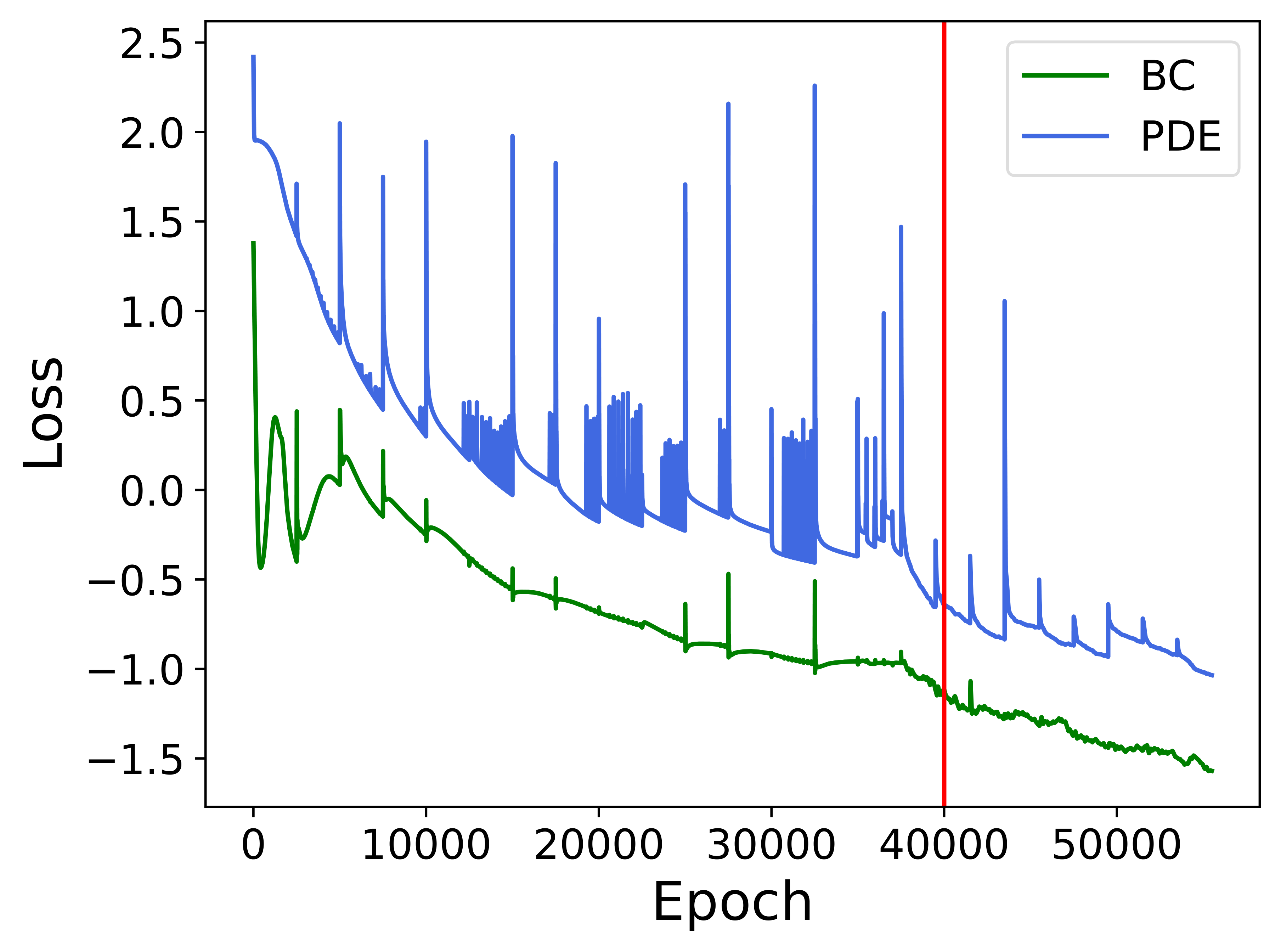}
\caption{Example of loss function behavior. The jumps correspond to the changes in boundary conditions and rearrangement of points used in training. The red vertical line marks the moment when the optimizer switches form \texttt{ADAM} to \texttt{LBFG-S}.}
\label{fig:losses}
\end{figure}

Another approach is to transform neural network output in a way that it satisfies boundary conditions automatically. For instance, a fulfillment of Dirichlet conditions on boundary $\partial D$ of a domain $D$ can be achieved if one introduces functions $f$ and $g$, such as $f|_{\mathbf{x} \in \partial D} = 0$ and $g|_{\mathbf{x} \in \partial D} = h_{\mathrm{BC}}$, where $h_{\mathrm{BC}}$ sets the boundary values. Then, function $g(\mathbf{x}) +f(\mathbf{x})  \mathcal{N}(\mathbf{x})$ will satisfy boundary conditions regardless of the function $\mathcal{N}(\mathbf{x})$.

In comparison with traditional solvers, the main advantages of the PINN method are the absence of a computational grid (which is important when dealing with complex domains) and lesser sensitivity to the dimensionality of the problem. The disadvantages are usually the lower accuracy and greater time expenses. 

It is also worth mentioning that the accuracy of most traditional numerical solvers tends to machine $\epsilon$ as computation time increases. At the same time, PINN's accuracy limit is determined by the neural network architecture and optimizer choice. Nevertheless, for many astrophysical problems (including the one considered in this paper) accuracy  is limited primarily by the choice of physical model and not by the numerical method itself. In such cases, PINN’s ability to solve equations with complex boundary conditions comes to the fore.

In application to the considered problem, PINN was used as follows. With a fixed vacuum gap height, $N\sim5000$ optimization steps are done; then, using the obtained potential, the gap height is recalculated and the procedure is repeated. It is important that the neural network weights obtained at the current iteration are used as the initial values of the weights at the next one, which helps to speed up the training process. As relative change in gap height becomes less than 1\%, the iterative procedure is stopped and additional training of the neural network is made with a final gap height.

We have chosen a fully-connected neural network, which consists of three hidden layers (20, 20, 20). In order to satisfy boundary conditions on $r_{\mathrm{m}} = 1$ and $z = 0$ surfaces, the output of the neural network was multiplied by the function $f(r_{\mathrm{m}}, z) = r_{\mathrm{m}} \cdot z$. At the same time, boundary condition on the upper surface of the gap was taken into account by adding an extra term into the loss function. For optimization, an approach proposed in a paper \citep{PINN_LBFG} was used.  It consists of primary training using \texttt{ADAM} optimizer \cite{ADAM} and subsequent training using \texttt{LBFG-S} method (see Fig. ~\ref{fig:losses}). While the \texttt{ADAM} method was applied directly in the iterative process, the \texttt{LBFG-S }method was used in the retraining with the final gap height. This approach is motivated by the fact that the \texttt{ADAM} method is much less susceptible to the problem of local minima, which is especially relevant at the initial stages of training, while the \texttt{LBFG-S} method allows for deeper optimization. Therefore, the strengths of both methods are combined and the weaknesses are leveled out, which made it possible to achieve significantly greater accuracy of calculations in the considered problem.

\subsection{Axisymmetric case}
\label{subsec:axisym_potential}
At inclination angles $\chi \lesssim 85 \text{\textdegree}$ the source term in the equation ~\eqref{eq:Poisson} is almost independent of the angle $\varphi_{\mathrm{m}}$. Thus, for non-orthogonal pulsars the problem becomes essentially axisymmetric and therefore 2-dimensional, which allows to exclude the $\varphi_{\mathrm{m}}$ angle from consideration. This significantly simplifies both the calculation and analysis of the results. In this regard, although the main focus of the study is the application of the above method to orthogonal pulsars, the axisymmetric case was considered first.
\begin{figure}[ht]
    \centering
    \begin{subfigure}[b]{0.95 \columnwidth}
       \centering 
       \includegraphics[width=\linewidth]{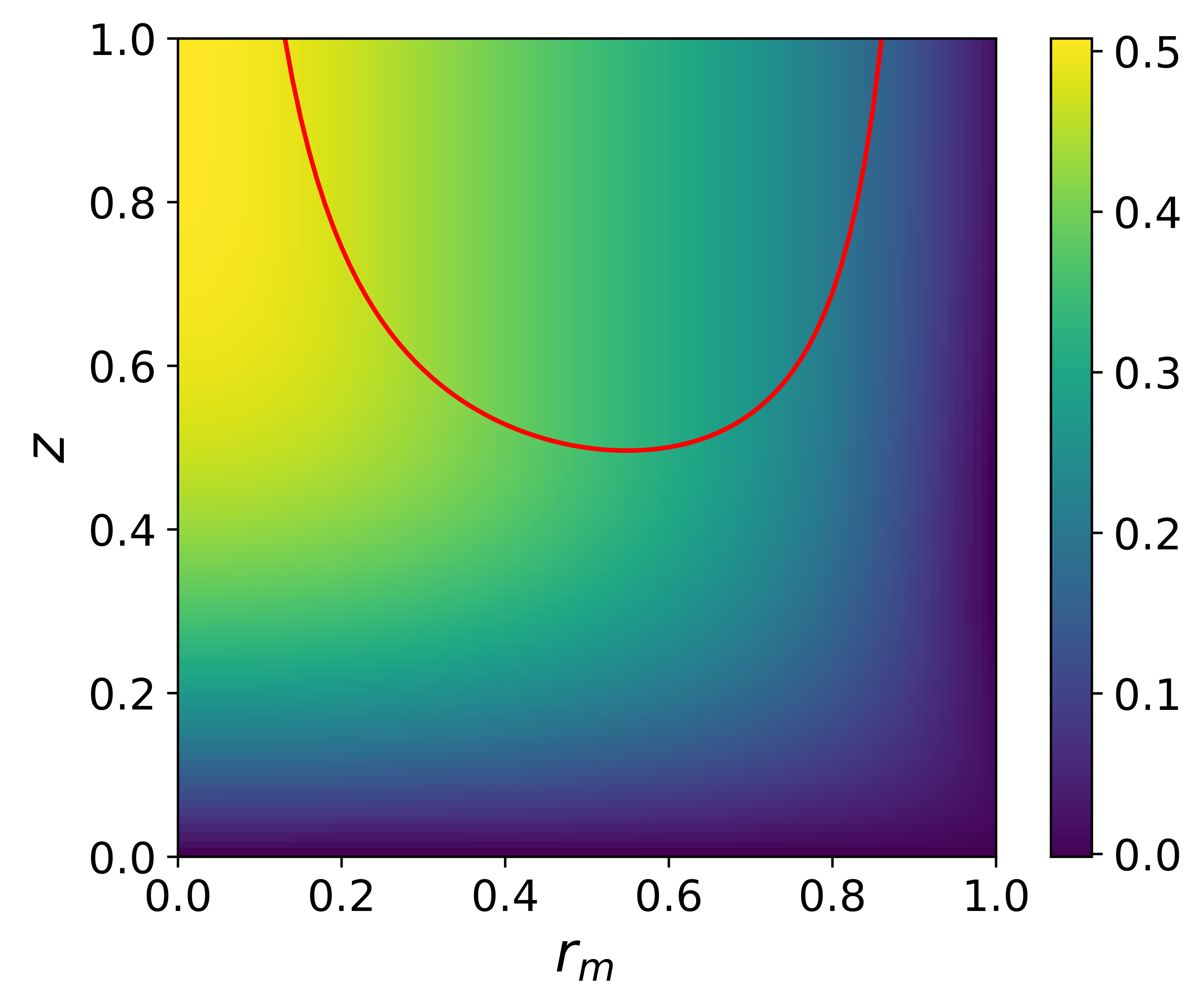}
       \caption{Potential, normalized to the quantity $\Omega B R_{0}^2/(2 c)$. The red line represents the gap height $H_{\mathrm{gap}}(r_{\mathrm{m}})$.}
    \end{subfigure}%
    \\
    \begin{subfigure}[b]{0.95 \columnwidth}
       \centering 
       \includegraphics[width=\linewidth]{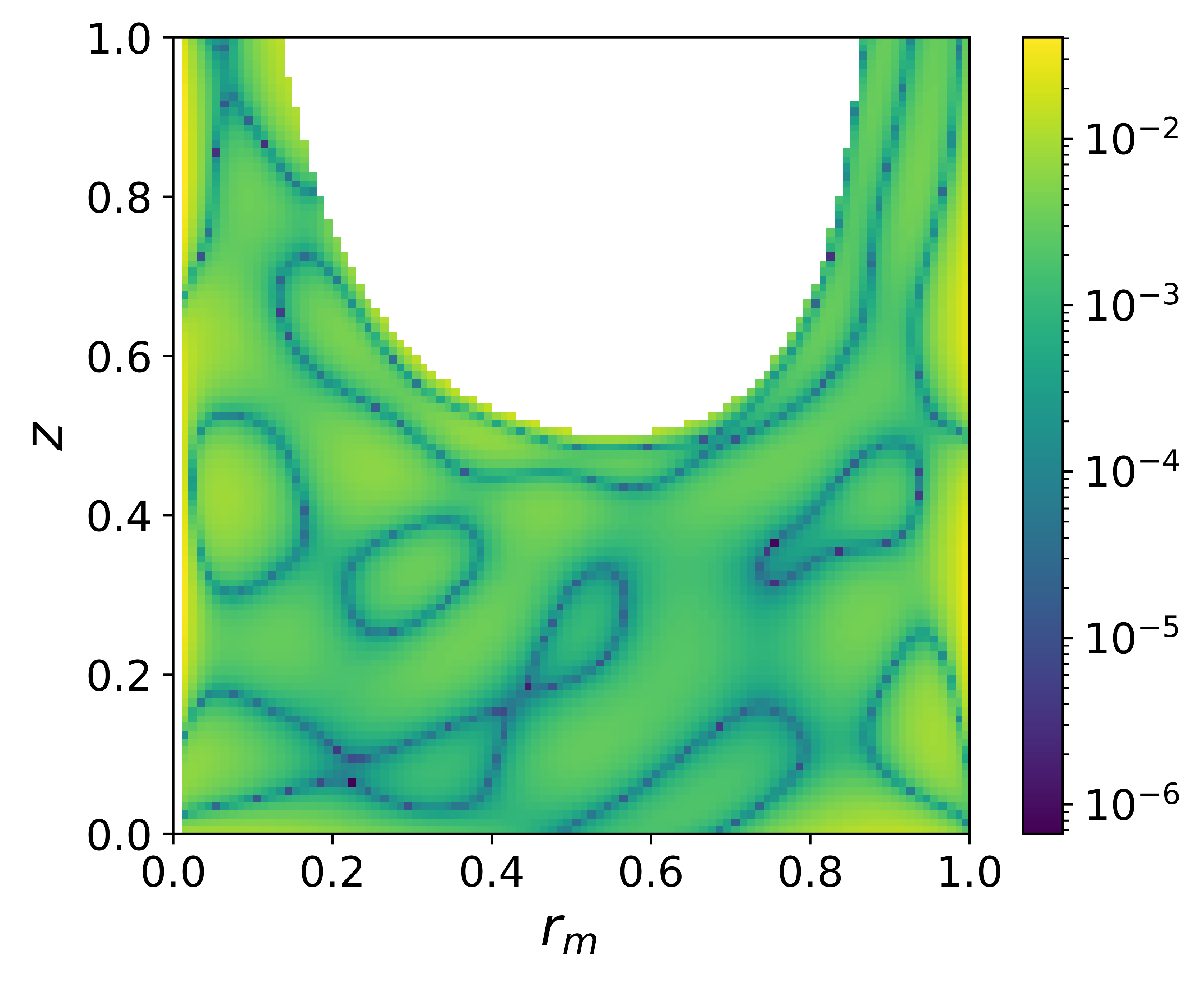}
       \caption{The norm of the residuals of Poisson equation}
    \end{subfigure}%
   \caption{Example of accelerating potential calculation for a pulsar with $\chi = 10 \degree$, $B_{12} = 1.6$ and $P = 0.5 \, \text{s}$.} 
   \label{fig:potential2D}
\end{figure}

\begin{figure*}[ht]
    \centering
    \begin{subfigure}[b]{0.3 \textwidth}
       \centering 
       \includegraphics[width=\textwidth]{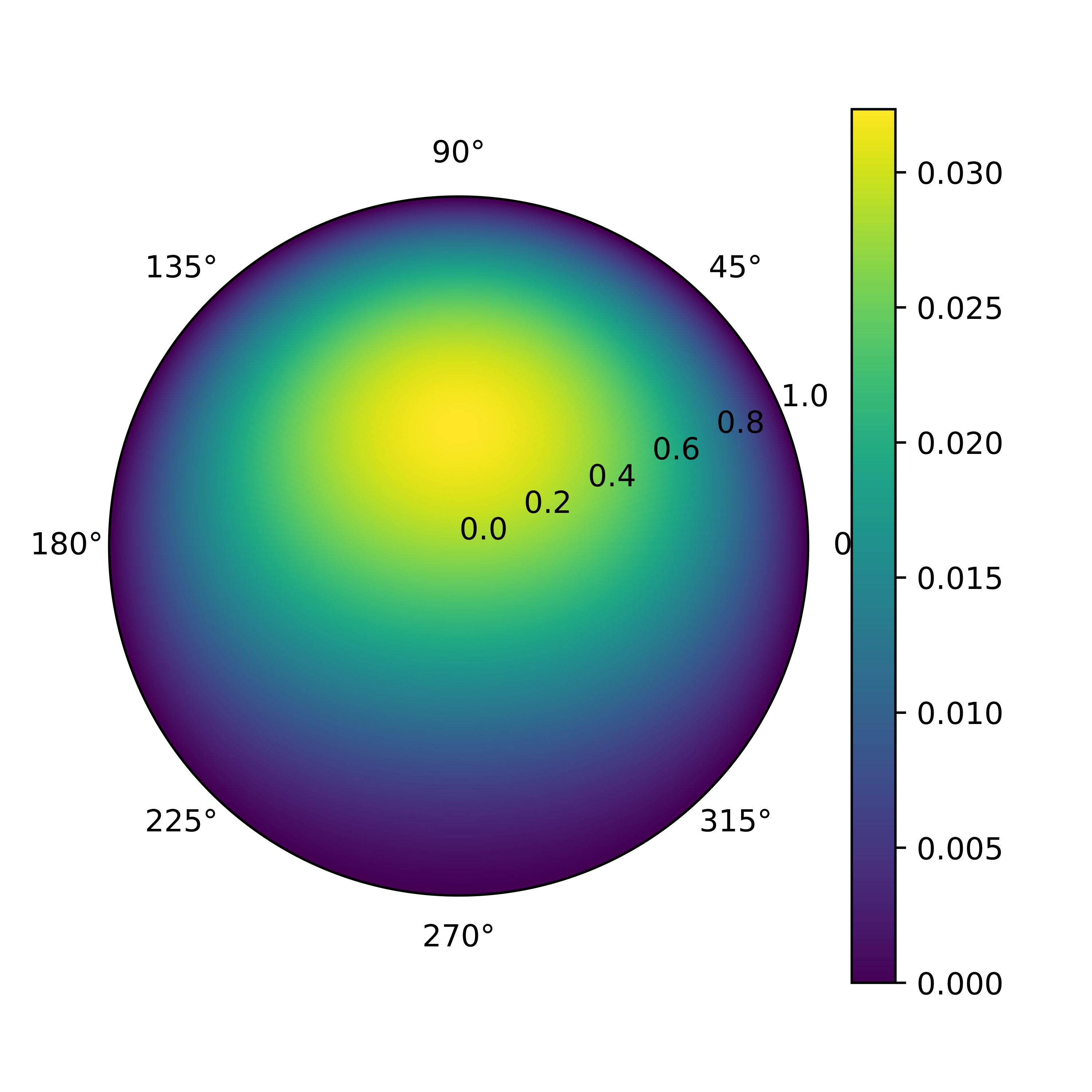}
    \end{subfigure}%
    \hspace{0.3cm}
    \begin{subfigure}[b]{0.3 \textwidth}
       \centering 
       \includegraphics[width=\textwidth]{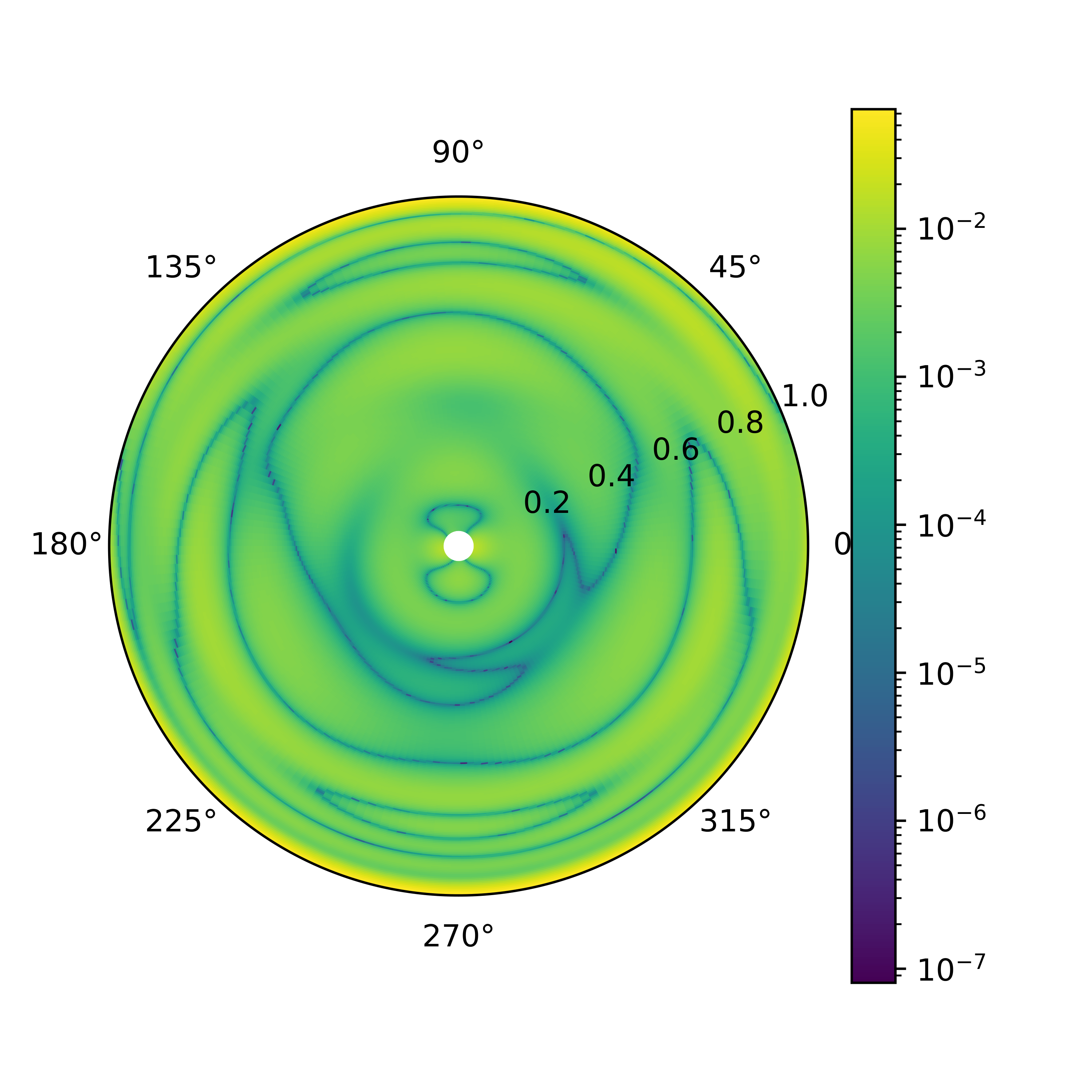}
    \end{subfigure}%
    \hspace{0.3cm}
    \begin{subfigure}[b]{0.3 \textwidth}
       \centering 
       \includegraphics[width=\textwidth]{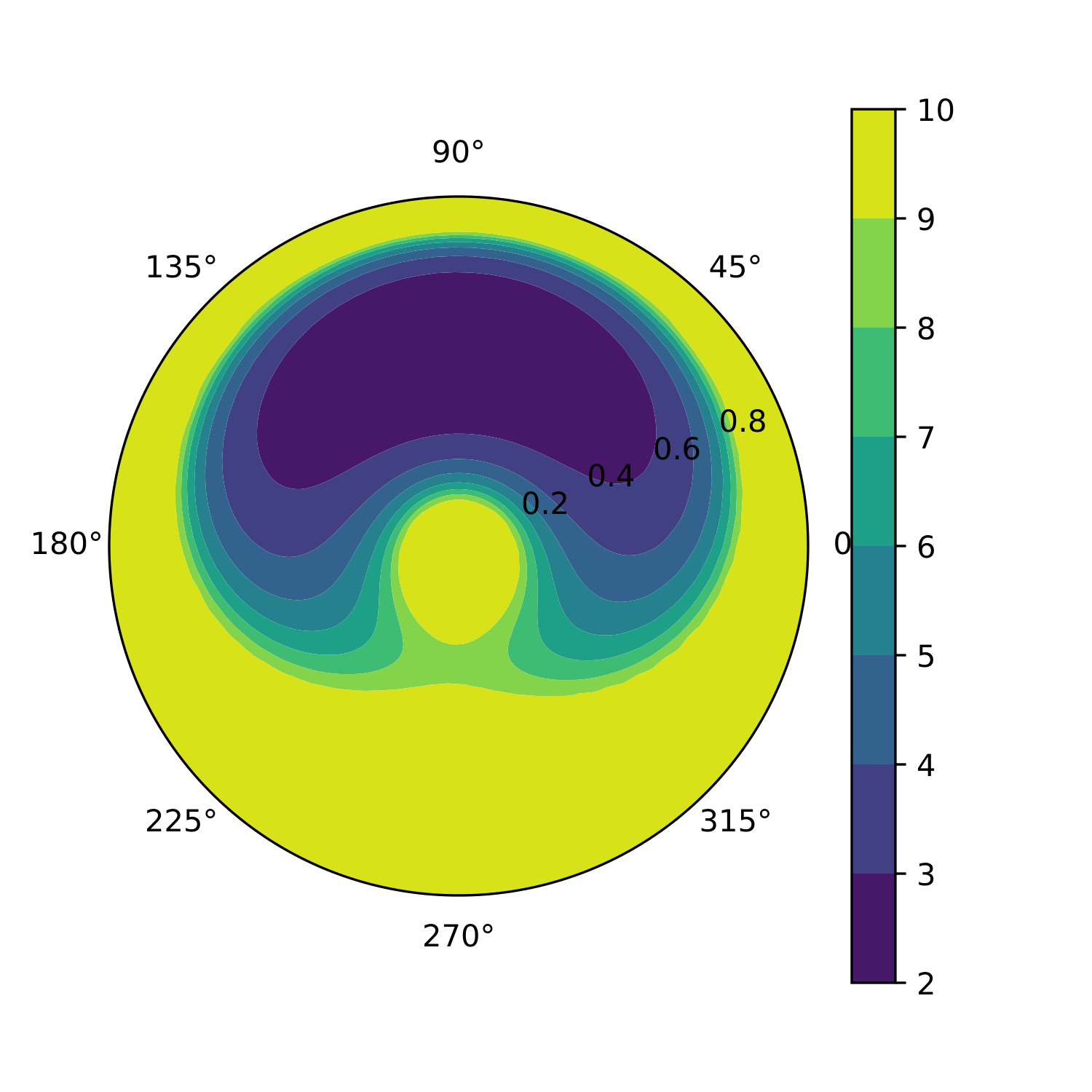}
    \end{subfigure}%

    \hspace{0.1cm}
        
    \begin{subfigure}[t]{0.3 \textwidth}
       \centering 
       \includegraphics[width=\textwidth]{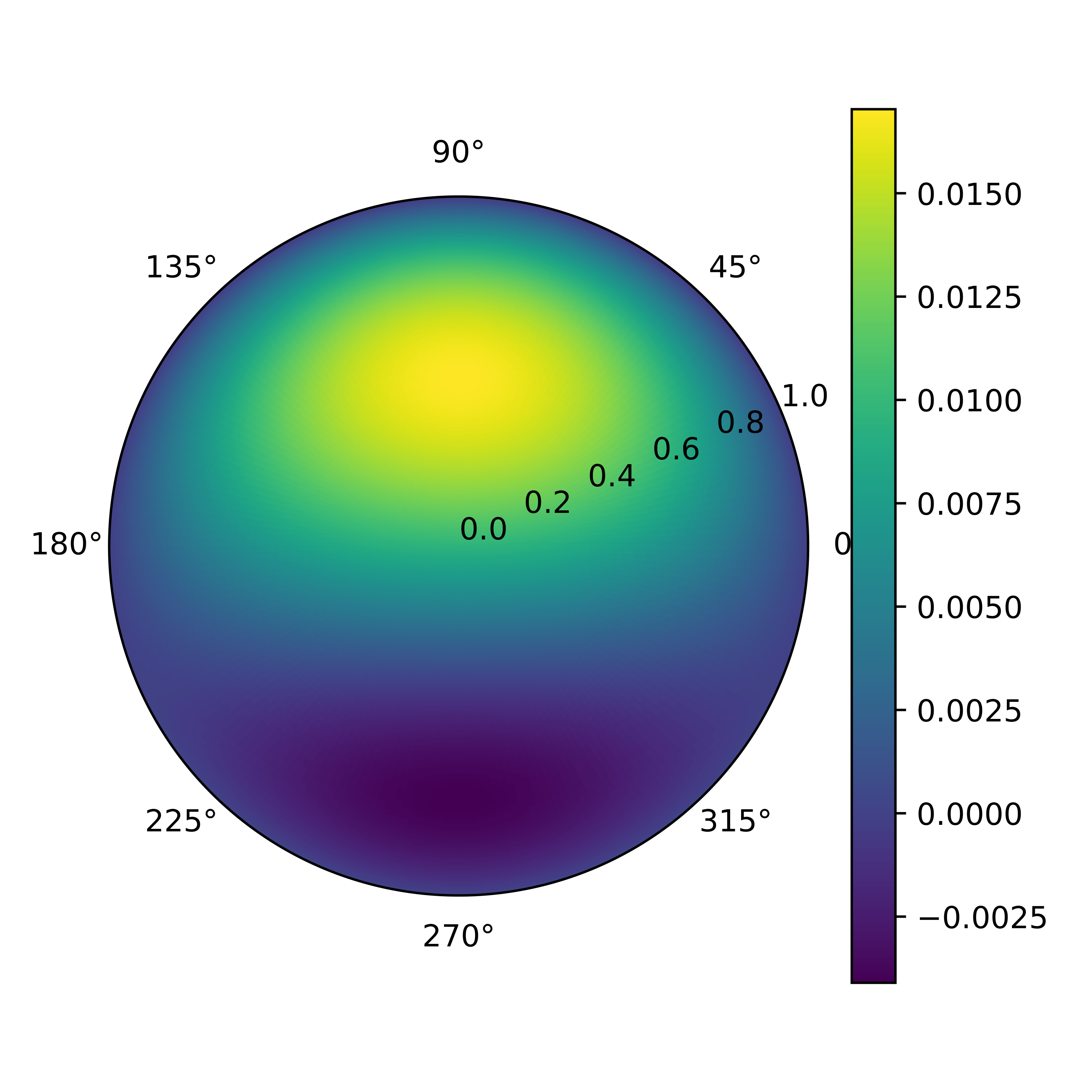}
       \caption{Accelerating potential, normalized to $\Omega B R_{0}^2/2c$} 
    \end{subfigure}%
    \hspace{0.3cm}
    \begin{subfigure}[t]{0.3 \textwidth}
       \centering 
       \includegraphics[width=\textwidth]{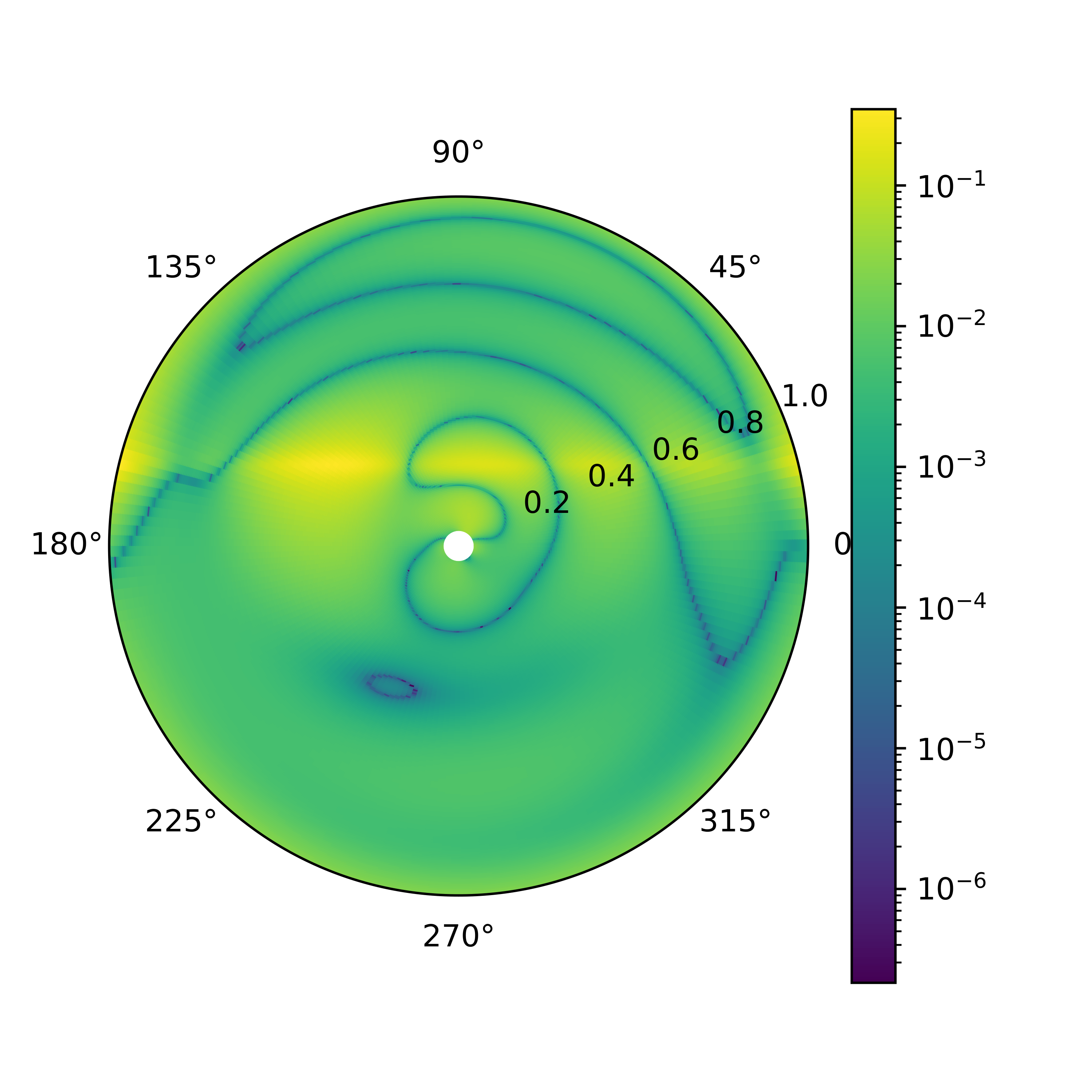}
       \caption{Solution relative error} 
    \end{subfigure}%
    \hspace{0.3cm}
    \begin{subfigure}[t]{0.3 \textwidth}
       \centering 
       \includegraphics[width=\textwidth]{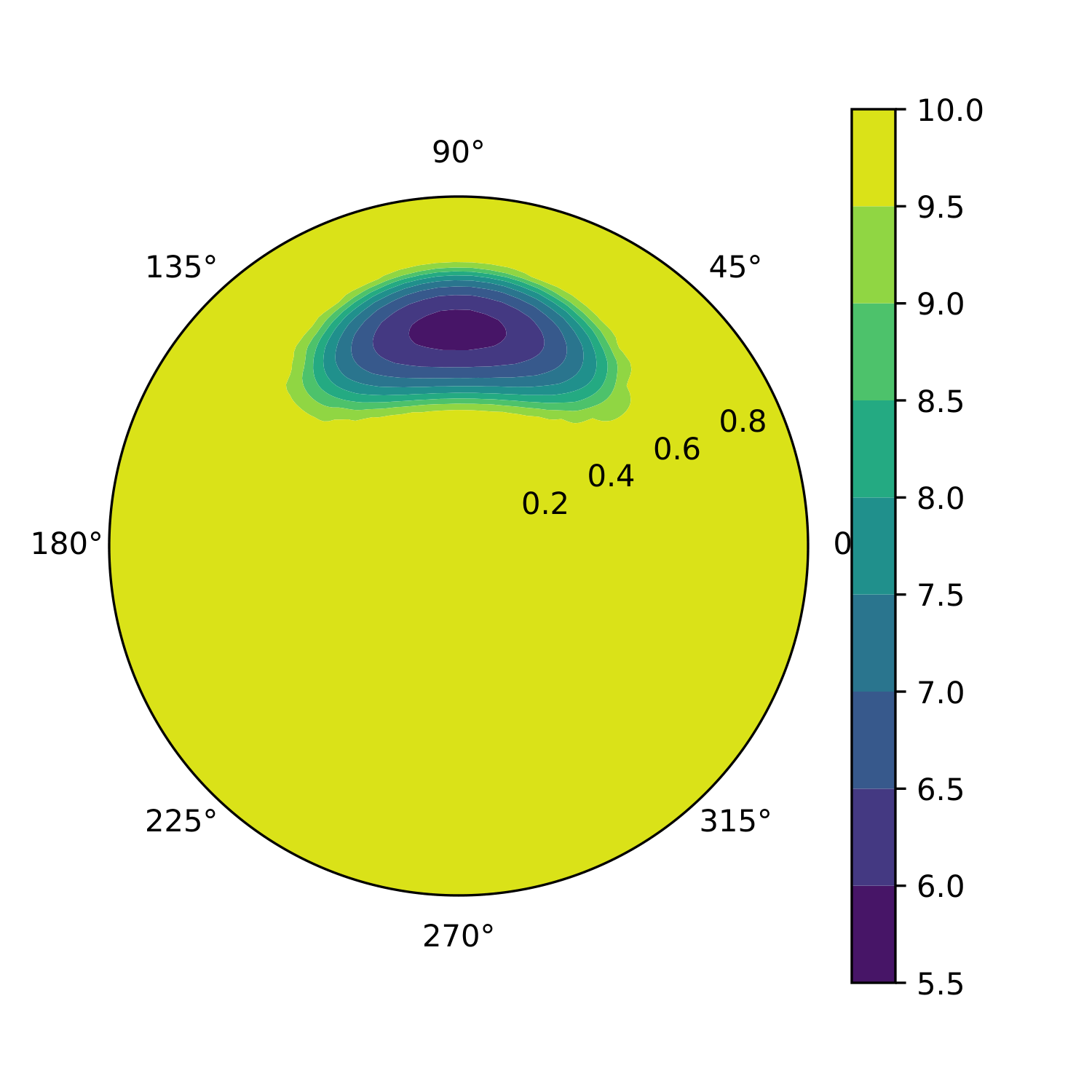}
       \caption{Vacuum gap height normalized to polar gap radius} 
    \end{subfigure}%
   \caption{Examples of accelerating potential calculation for orthogonal pulsars with $B_{12} = 1.5$, $P = 0.3$ s, $\chi= 88\degree$ for the upper row and $\chi = 89.3 \degree$ --- for the lower one. From left to right accelerating potentials magnitudes, Poisson equation relative errors on the $0.5 R_0 $ cut and gap heights normalized to polar cap radius are shown. } 
   \label{fig:potentials3D}
\end{figure*}

\begin{figure}[ht]
\centering
\includegraphics[width=0.95\linewidth]{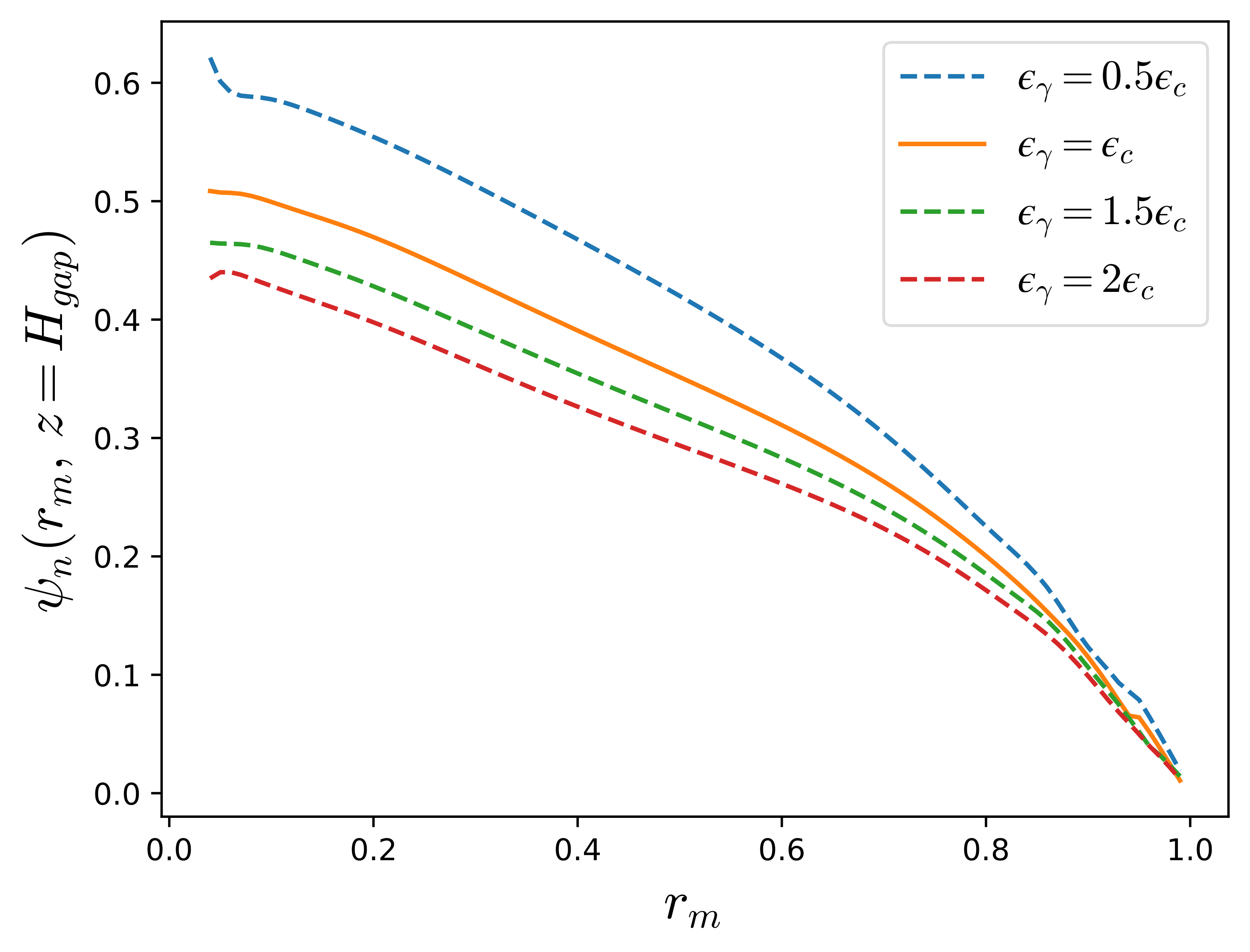}
\caption{Accelerating potential profiles at the gap height, calculated for different characteristic energies of curvature radiation for a pulsar with  $\chi = 10 \degree$, $B_{12} = 1.6$ and $P = 0.5 \, \text{s}$. Potential is normalized to the quantity $\Omega B R_{0}^2/2c$.}
\label{fig:psi_max}
\end{figure}

As one can see in Fig.~\ref{fig:potential2D}, the proposed method allowed us to find a solution with a good enough accuracy of an order of $(1-2)\%$. Moreover, obtained results reproduce some geometrical properties of the structure, known as ''slot-gap'' \citep{SlotGap}. However, it is worth mentioning that in the work ~\citep{SlotGap} a stationary model with a free particle escape was considered, which means that $\rho_{e}$ and $\rho_{\rm GJ}$ charge densities in  (\ref{eq:PoissonGeneral}) are close to each other. Nevertheless, in both cases, this shape of the gap is due to the fact that near the magnetic axis, the generation of particles is suppressed by the small curvature of the magnetic field, and at the edge of the polar cap, the accelerating potential should tend to zero. Therefore, the gap height must tend to infinity at  $r_m \sim 0$ and $r_m \sim 1$. 
The Fig. \ref{fig:psi_max} shows limiting values of accelerating potential (achieved at the height of the vacuum gap) which were calculated using different characteristic energies in the monoenergetic approximation of the synchrotron spectrum \eqref{eq:mono_spectrum}. As one can see, even when the characteristic energy of a photon changes twofold,  the relative change in the potential does not exceed $\sim (15-20) \% $, which is an acceptable error within the framework of this study.  However, the effects of the curvature spectrum will be considered in subsequent studies.

\subsection{Non-Axisymmetric case}

On the Fig.~\ref{fig:potentials3D} one can see examples of accelerating potential calculation for orthogonal pulsars with inclination angles $\chi= 88\degree$ (upper row) and  $\chi = 89.3 \degree$ (lower row), magnetic field $B_{12} = 1.5$ and period $P = 0.3$ s, typical for orthogonal interpulse pulsars. The first column shows the limiting values of accelerating potential along the field lines, the second column shows cuts of the Poisson equation residuals at the height of $0.5 R_0 $, and in the third one, vacuum gap heights normalized to the polar cap radius are presented. It is worth mentioning that the height $H_{\mathrm{gap}} = 10 R_0$, in fact, corresponds to infinite gap height $H_{\mathrm{gap}} \rightarrow \infty$. 

As was already mentioned, the study of orthogonal pulsars is of particular interest, since the electric potential becomes significantly non-axisymmetric for such objects. Moreover, in this case, the Goldreich---Julian charge density can change sign within the polar cap, which entails a change in the direction of the accelerating field. However, this occurs in a very narrow range of inclination angles $\chi$.

Indeed, in order for $\cos \theta_{\mathrm{b}}$ to change its sign within the polar cap, it is necessary that $|\chi - \pi/2| \lesssim 3R_{0}/2R \sim 0.03$ (here we assumed for the period to be $P \approx 1$). Accordingly, for the angle $\chi$ we obtain  $88.5\degree \lesssim \chi \lesssim 91.5\degree$. As a result, as can be seen on Fig. ~\ref{fig:potentials3D}, the change in the sign of the potential takes place only for the angle $\chi = 89.3 \degree$, while for the angle $\chi= 88\degree$ the sign of the potential does not change within the polar cap.

We finally note that the magnitude of accelerating potential appears to be significantly smaller than for non-orthogonal pulsars. This is directly related to the small value of $\cos{\theta_{\rm b}}$ in the Goldreich---Julian charge density \eqref{eq:rho_GJ}. 

\section{Secondary plasma density profile}
\label{sec:Dencity}

\subsection{Calculation method}
To study the spatial distribution of plasma density, it is convenient to write down the plasma number density in the following form: $n = \lambda g(r_{\mathrm{m}}, \varphi_{\mathrm{m}} ) n_{\mathrm{GJ}}$, where $n_{\mathrm{GJ}} = |\rho_{\mathrm{GJ}}| / e $ is the Goldreich-Julian number density, $\lambda$ is the multiplicity parameter, determined by averaging the value $n / n_{\mathrm{GJ}}$ over the polar cap and the $g(r_{\mathrm{m}}, \varphi_{\mathrm{m}})$ is a dimensionless factor, constant on the magnetic field lines.  In other words, the entire relationship between the plasma density and the height above the polar cap is encapsulated in the Goldreich-Julian number density $n_{\mathrm{GJ}}$. 
However, since the value of the $n_{\mathrm{GJ}}$ itself depends significantly on the coordinates $r_{\mathrm{m}}$, $\varphi_{\mathrm{m}}$ for orthogonal pulsars, we have chosen the following value as the normalization
\begin{equation}
    n_0 = \frac{\Omega B}{2 \pi c e} \sqrt{\frac{\Omega R}{c}},
    \label{eq:dencity_nomralization}
\end{equation}
where the last multiplier corresponds to the characteristic value of $\cos{\theta_{\mathrm{b}}}$ on the polar cap of an orthogonal pulsar. Then, the multiplicity parameter $\lambda$ describes the efficiency of the pair creation and the factor $g(r_{\mathrm{m}}, \phi_{\mathrm{m}} ) $ describes its spatial distribution.

To calculate the plasma multiplicity, we used the following model: primary particles, accelerated to the values of the Lorentz factors $\gamma_{\mathrm{e}} \sim 10^7$ in the potential $\psi(r_{\mathrm{m}}, \varphi_{\mathrm{m}}, z)$, obtained in the first part of our work, emit primary photons, which are then absorbed in a magnetic field and give rise to a synchrotron cascade of particle production \citep{Omode, HibAronsPairProduction}. Considering that the free path of photons is much shorter than the distance from the star center to the radiation point, we can assume that the synchrotron cascade is local in space. Based on this assumption, we derive an expression for the number of secondary particles produced during the production of one primary particle within the framework of this model.
\begin{equation}
\lambda_1 = \int_{0}^{\infty} \mathrm{d} z_e \int_{0}^{\infty} \mathrm{d} \epsilon_i n_{\gamma} (z_e, \epsilon_i; r_{\mathrm{m}}, \varphi_{\mathrm{m}}) f_{\mathrm{SR}} (z_e, \epsilon_i; r_{\mathrm{m}}).
\end{equation}  
Here $z_e$ is the height of the radiation of the curvature photon and  $n_{\gamma}$ --- is the spectrum of primary photons. Finally, the function $f_{\mathrm{SR}}$ determines the number of secondary particles produced by a single photon with energy $\epsilon_i$, that is absorbed at a height $z_a$. To obtain the $f_{\mathrm{SR}}$ function, we decided to use the method proposed in the paper \citep{HibAronsPairProduction}, but this time the calculations will take into account the dependence of all quantities on coordinates on the polar cap.

\begin{figure}[ht]
\centering
\includegraphics[width=0.9\linewidth]{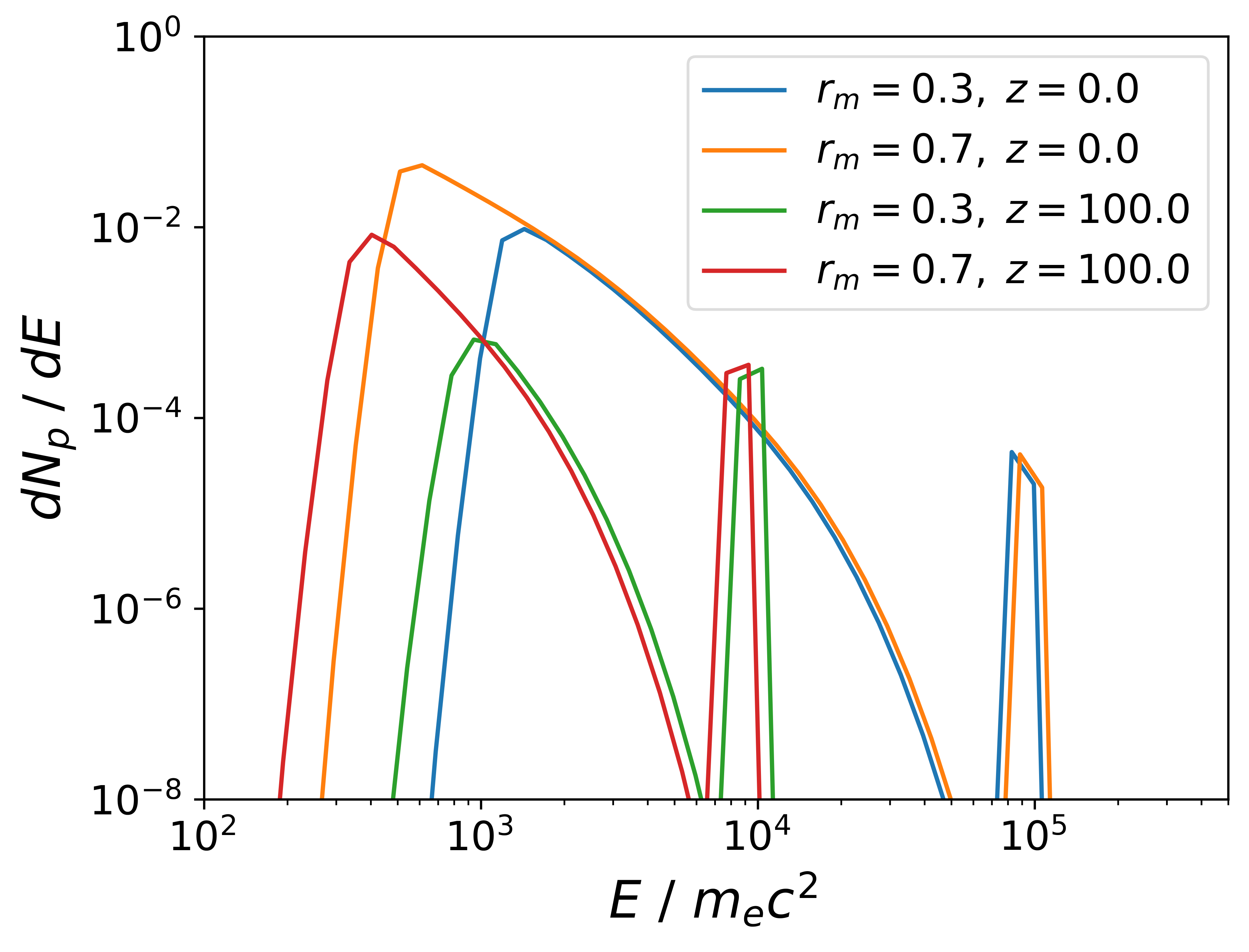}
\caption{Spectrum of a synchrotron cascade triggered by a photon with energy $\epsilon_{\gamma} = 10^6$ for different values $r_{\rm m}$ and $z$.}
\label{fig:OTSSpectra}
\end{figure}

If we introduce the local volumetric rate of pair $Q_{\mathrm{p}}$ and photon $Q_{\gamma}$ production events, we can write the following expression for the number of secondary particles $f_{\mathrm{SR}}$:
\begin{equation}
    f_{\mathrm{SR}} (\epsilon_i) = \int_0^{\infty} \mathrm{d} \epsilon Q_{\mathrm{p}} (\epsilon, \epsilon_i) = \int_0^{\infty} \mathrm{d} \epsilon (1 - e^{-\tau_{\infty}(\epsilon)}) Q_{\gamma} (\epsilon, \epsilon_i).
\end{equation}
Here $\tau_{\infty}$ is the limiting optical opacity of the magnetosphere. In turn $Q_{\gamma}$ can be determined from the integral equation \citep{HibAronsPairProduction}:
\begin{equation}
    \begin{aligned}
    Q_{\gamma}(\epsilon, \epsilon_i)= & \int_0^{\infty} d \epsilon'\left[1-e^{-\tau_{\infty}\left(\epsilon'\right)}\right] \\
    & \times \frac{1}{\epsilon'} K\left(\frac{\epsilon}{\epsilon'}\right)\left[  \delta(\epsilon' - \epsilon_i) +Q_{\gamma}\left(\epsilon', \epsilon \right)\right].
    \end{aligned}
    \label{eq:integral_equation}
\end{equation}
In this equation, function $K(\epsilon / \epsilon_i)$ has the form
\begin{equation}
    \begin{aligned}
K\left(\frac{\epsilon}{\epsilon'}\right)= & \frac{3 \sqrt{3}}{8 \pi} \sqrt{\Lambda}\left(\frac{\epsilon}{\epsilon'}\right)^{-3 / 2} \\
& \times\left[G\left(\Lambda \frac{\epsilon}{\epsilon'}\right)-G\left( (1 + a^2) \Lambda \frac{\epsilon}{\epsilon'}\right)\right],
    \end{aligned}
\end{equation}
where $a =  3  B_{\mathrm{cr}} /(4 \ln \Lambda B)$ and $\Lambda$ is determined by
expression \eqref{eq:Lambda}. In turn,
\begin{equation}
    G(t) = \int^{\infty}_{t} {\rm d}x K_{5/3}(x) (x^{3/2} - t^{3/2}).
\end{equation}
The solution of this equation can be easily obtained by introducing discretization in logarithmic scale and solving the corresponding matrix equation. The Fig. \ref{fig:OTSSpectra} shows examples of calculations of the secondary particle spectrum, depending on the location of cascade formation. As we can see, as the height $z$ increases, a power-law distribution of secondary particles forms. This well-established fact ~\citep{Omode, HibAronsPairProduction}  confirms the accuracy of our analysis.

Furthermore, as well as when calculating the potential, to determine the spectrum of the primary photons  $n_{\gamma}$, it is necessary to take into account both curvature radiation and inverse Compton scattering. To estimate the contribution of Compton scattering, approximate expressions can be used for the rate of pair production (the number of photons produced by one primary particle)\citep{HibAronsPulsarDeath}:
\begin{equation}
    \dot N_{NR} \approx 1.5 \cdot 10^9 \gamma^{-1} T_6^2 \Delta \mu \: \text{s}^{-1},
\end{equation}
\begin{equation}
    \dot N_{R} \approx 10^{13} \gamma^{-2} B_{12} T_6 \: \text{s}^{-1},
\end{equation}
\begin{equation}
    \dot N_{\mathrm{CR}} \approx \gamma \rho_8^{-1} \: \text{s}^{-1}.
\end{equation}
Here, the indices $NR$, $R$ and $\mathrm{CR}$ refer to non-resonant Compton scattering in the Klein-Nishina limit, resonant Compton scattering, and curvature radiation. Here $T_6$ is the polar cap temperature in units of $10^6$ K, $\rho_8$ is the radius of curvature of magnetic field lines in units \mbox{$10^8$ cm,} and $\Delta \mu = 1 - \mu_{\mathrm{min}}$.

It is possible to compare the efficiency of secondary plasma generation for the three mechanisms by multiplying the values  $\dot{N}$ by a factor $f_{\mathrm{SR}}(\epsilon_{\gamma}(\gamma_{\mathrm{e}}))$, where the mechanism of primary photon generation determines the form of $\epsilon_{\gamma}(\gamma_{\mathrm{e}})$. For example, for curvature radiation, $\epsilon_{\gamma} \sim 3 / 2 \; \lambda_e / R_c \; \gamma_e^3$, for non-resonant inverse Compton scattering in the Klein-Nishina limit $\epsilon_{\gamma} \sim \gamma_e$, and for
resonant --- $\epsilon_{\gamma} \sim 2 \epsilon_B \gamma_e$. 
Considering that the characteristic scale of secondary plasma generation is the scale of magnetic field decay, which is the radius of the star, we can estimate the value $\lambda_1$ as $\dot{N}(\gamma_{\mathrm{e}}) f_{SR} R /c$.

\begin{figure}[ht]
\centering
\includegraphics[width=0.9\linewidth]{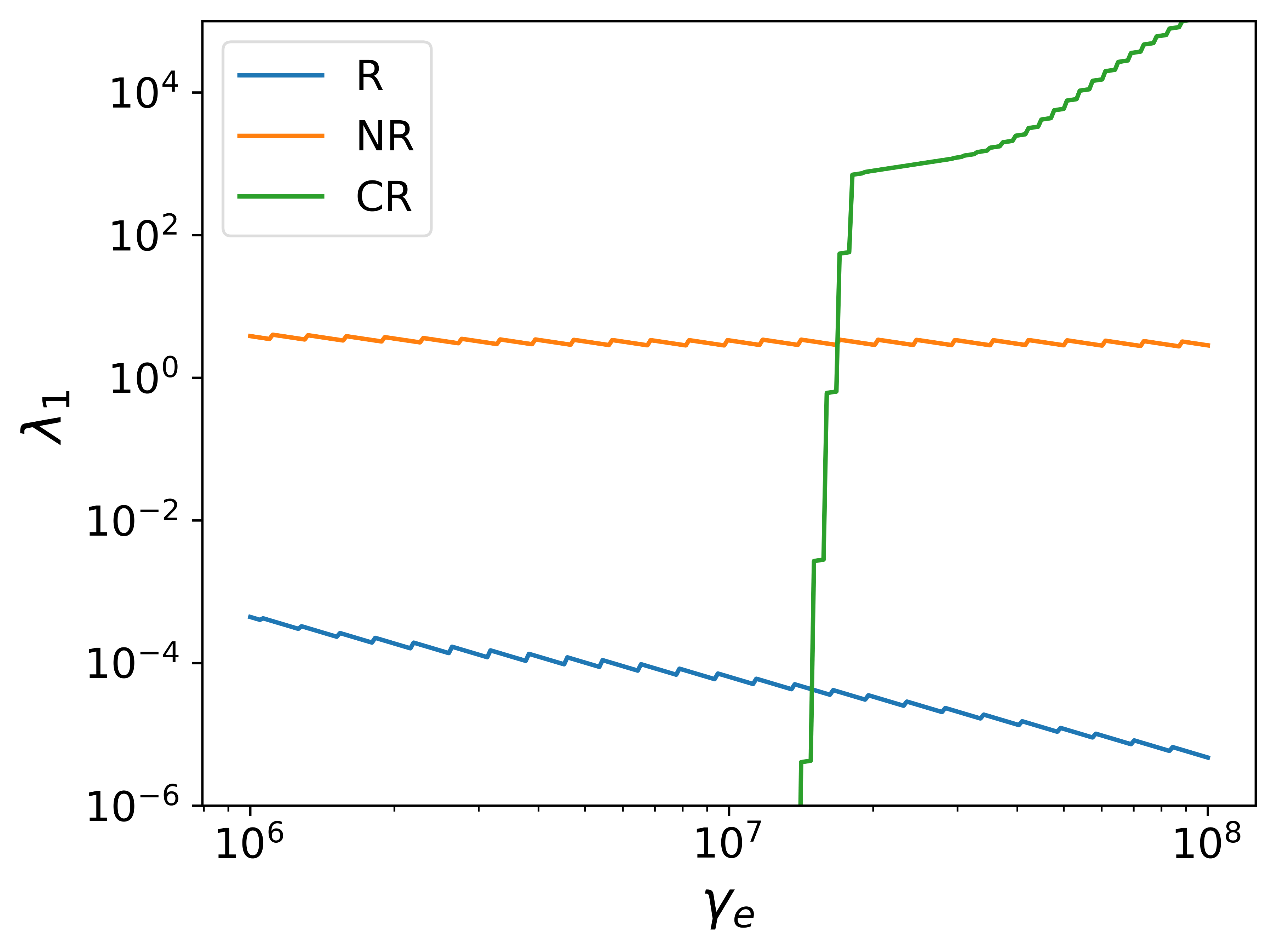}
\caption{Comparison of the efficiency of secondary plasma generation for curvature radiation, inverse resonant, and non-resonant Compton scattering.}
\label{fig:MNdot}
\end{figure}

The results of the calculations for the $\gamma$-factors of the primary particles of interest are shown in the Fig.~\ref{fig:MNdot} As you can see, inverse Compton scattering can be ignored in the entire range of energies that we are interested in. Additionally, the results for non-resonant scattering, shown in the Fig.~\ref{fig:MNdot}, were obtained for a $\Delta \mu = 1$, while characteristic $\Delta \mu$ values at altitudes around the star's radius are around $\left(R_0/R\right)^2 \ll 1$.

As mentioned above, since the scale of secondary plasma generation is several star radii, obtained results are significantly overestimated. Therefore, in this section, we will limit ourselves to considering only the curvature mechanism. In this case, the number density of primary photons $n_{\gamma}$ is determined by the curvature radiation spectrum:

\begin{equation}
    \mathrm{d} N_{\gamma}^{(1)}=\frac{\sqrt{3}}{2 \pi} \frac{e^2}{c R_{\mathrm{c}}(z)} \frac{\gamma_{\mathrm{e}} F\left(\omega / \omega_{\mathrm{c}}\right)}{\hbar \omega} \mathrm{d} \omega \mathrm{d} z \equiv n_{\gamma} \mathrm{d} \omega \mathrm{d} z,
    \label{curvature_spectrum}
\end{equation}
where
\begin{equation}
    F(t)=t \int_{t}^{\infty} K_{5 / 3}(x) \mathrm{d} x
\end{equation}
and
\begin{equation}
    \omega_c = \frac{3}{2}\frac{c}{R_{\mathrm{c}}}\gamma_{\mathrm{e}}^3.
\end{equation}

Once we determine the number of particles produced by one primary particle $\lambda_1$, we can find the number density of secondary plasma by multiplying the $\lambda_1$ by the density of primary particles, which is equal to the Goldreich-Julian number density. Then, according to the definition \eqref{eq:dencity_nomralization}, we have 
\begin{equation}
    \lambda g(r_{\mathrm{m}}, \phi_{\mathrm{m}}) = \lambda_1 (r_{\mathrm{m}}, \phi_{\mathrm{m}}) \cdot
    \begin{cases}
    1, & \chi < 85 \degree, \\
    \cos{\theta_{\mathrm{b}}}/\sqrt{\frac{\Omega R}{c}}, & \chi > 85 \degree.
    \end{cases}
\end{equation}

\subsection{Axisymmetric case}

Same as for the accelerating potential, it is convenient to consider non-orthogonal and orthogonal pulsars separately. In the axisymmetric case, it is interesting to compare the results with the work \citep{Omode}, where similar calculations were carried out for pulsars close to the death line (i.e. the entire magnetosphere was assumed to be vacuum). Although the qualitative form of the profiles obtained coincides with the results of this work, there are several significant differences.

\begin{figure}[ht]
\centering
\includegraphics[width=0.9\linewidth]{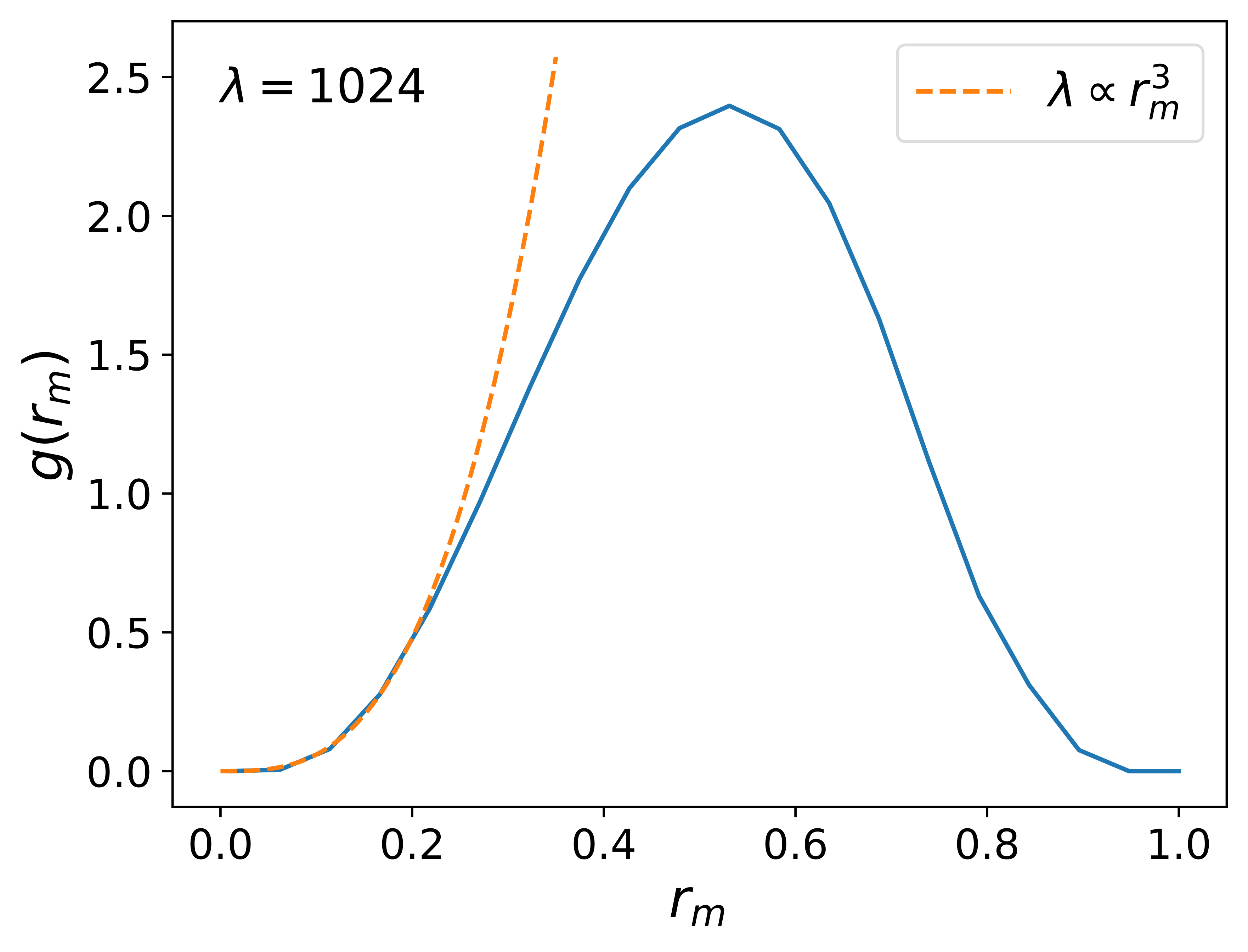}
\caption{Secondary plasma concentration profile $g(r_{\mathrm{m}})$ for a pulsar with parameters $P = 0.5 \, \text{c}$, \mbox{$B_{12} = 1.0$,} $\chi = 10\degree$. The dashed line shows the approximation of $g \propto r_{\rm m}^3$.}
\label{ns2D}
\end{figure}

\begin{figure}[ht]
\centering
\includegraphics[width=0.9\linewidth]{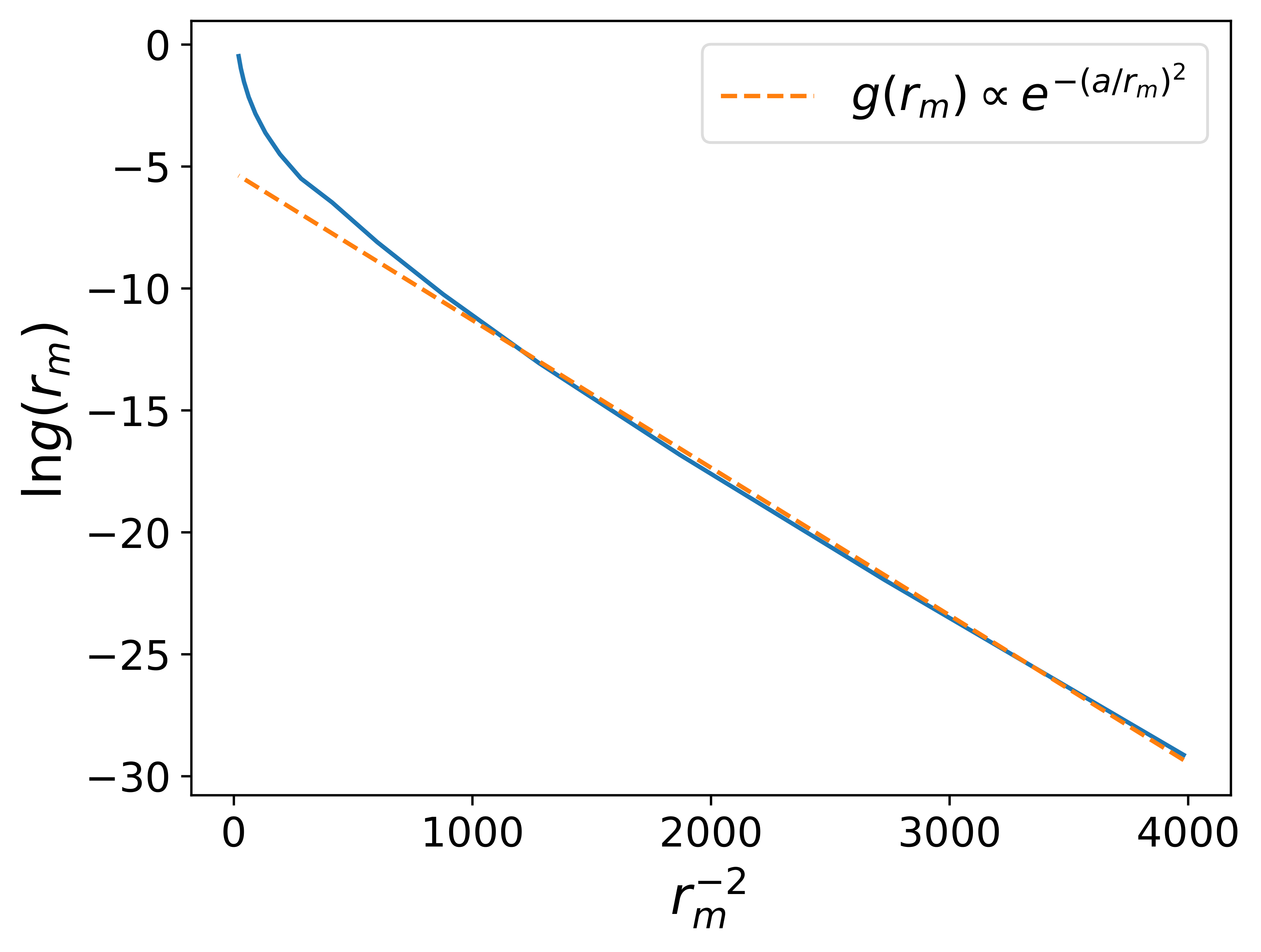}
\caption{Linearization of the behavior of the concentration profile in the limit of $r_{\mathrm{m}} \rightarrow 0$. The dashed line shows the asymptotics of $g(r_m) \propto e^{-\left(a/r_{\rm m}\right)^2}$.}
\label{ns_asympt}
\end{figure}

First of all, it has been found that the energy of the primary photons is typically not sufficient to initiate a multi-stage synchrotron cascade. In other words, the value of $f_{\mathrm{SR}}$ is close to 1 for the curvature photon energies. Accordingly, the multiplicity of secondary plasma with the same parameters turned out to be less than in the work \citep{Omode}.  Another difference is the asymptotic behavior of the number density profile near the magnetic axis. While in ~\citep{Omode} based on qualitative considerations, it was argued that with a good accuracy, it can be set  $g(r_{\mathrm{m}}) \propto r_{\mathrm{m}}^3$ at $r_{\mathrm{m}} \ll 1$, in this study, the result of $g(r_{\mathrm{m}}) \propto \exp[-a^2/r_{\mathrm{m}}^2]$, $a \ll 1$ was obtained (see Fig.~\ref{ns2D}-\ref{ns_asympt}). This asymptotic behavior follows from the expansion of the curvature radiation spectrum \eqref{curvature_spectrum} near the magnetic axis, which was not taken into account in the work ~\citep{Omode}. However, it is worth noting that the above asymptotics is applicable only near the magnetic axis ($r_\mathrm{m}  < 0.03$), while the cubic dependence better describes the profile on a larger scale $0.03 < r_\mathrm{m}  < 0.2$. For practical purposes, it is recommended to use asymptotics $g(r_{\mathrm{m}}) \propto r_{\mathrm{m}}^3$.

\begin{figure*}
    \centering
    \begin{subfigure}[b]{0.3 \textwidth}
       \centering 
       \includegraphics[width=\textwidth]{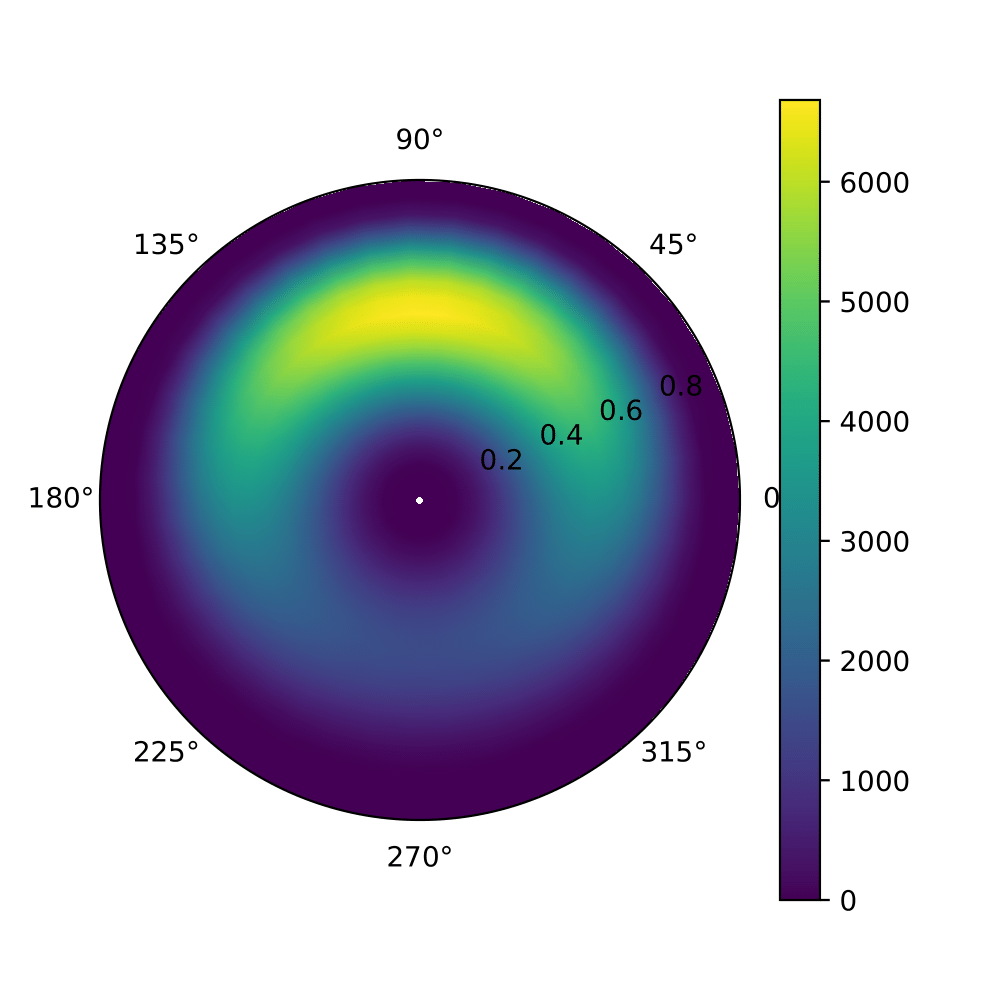}
       \caption{$P = 0.3\text{ c}$, $B_{12} = {1.5}$, $\chi = 85 \degree$.} 
    \end{subfigure}%
    \hspace{0.3cm}
    \begin{subfigure}[b]{0.3 \textwidth}
       \centering 
       \includegraphics[width=\textwidth]{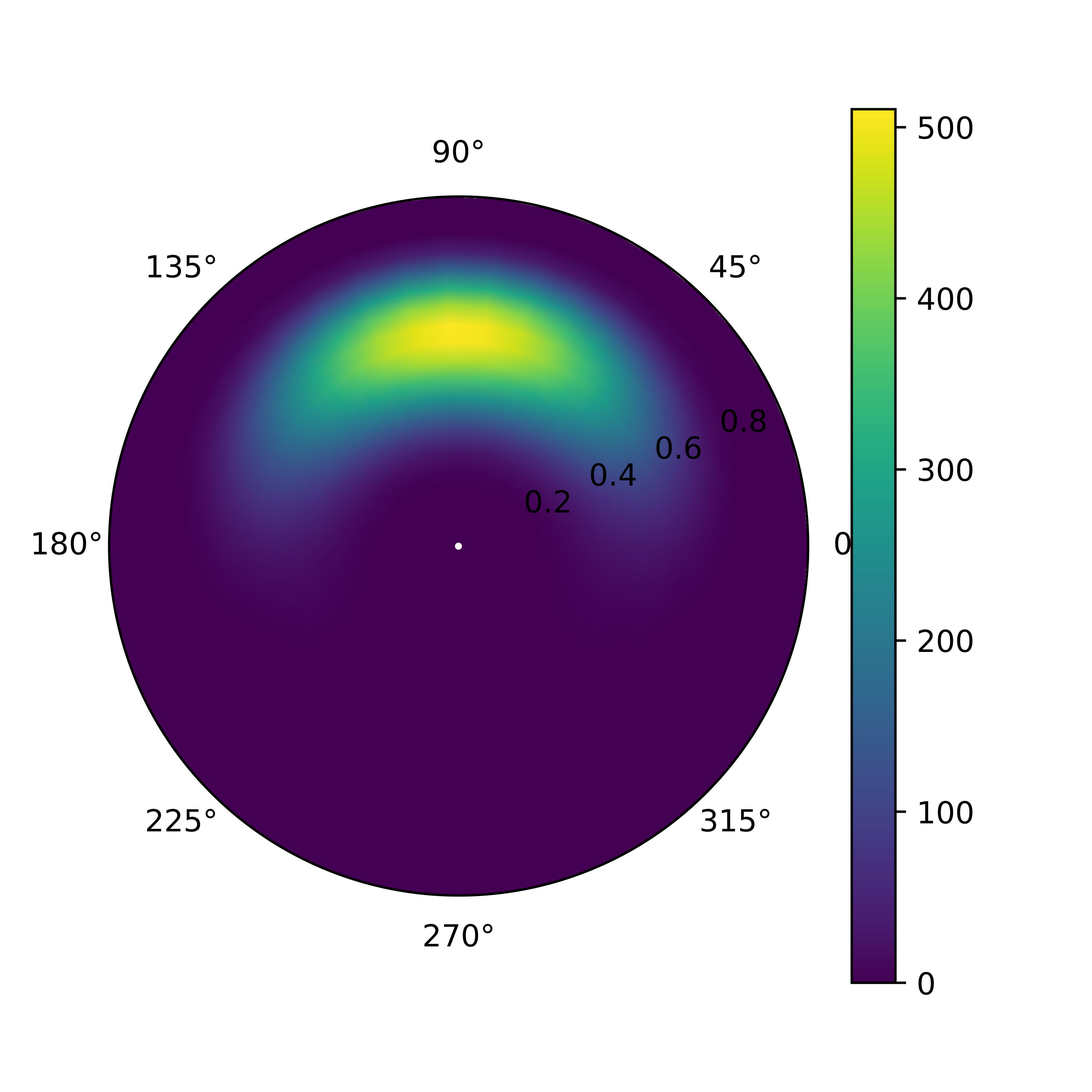}
       \caption{$P = 0.3\text{ c}$, $B_{12} = {1.5}$ $\chi = 88 \degree$.} 
    \end{subfigure}%
    \hspace{0.3cm}
    \begin{subfigure}[b]{0.3 \textwidth}
       \centering 
       \includegraphics[width=\textwidth]{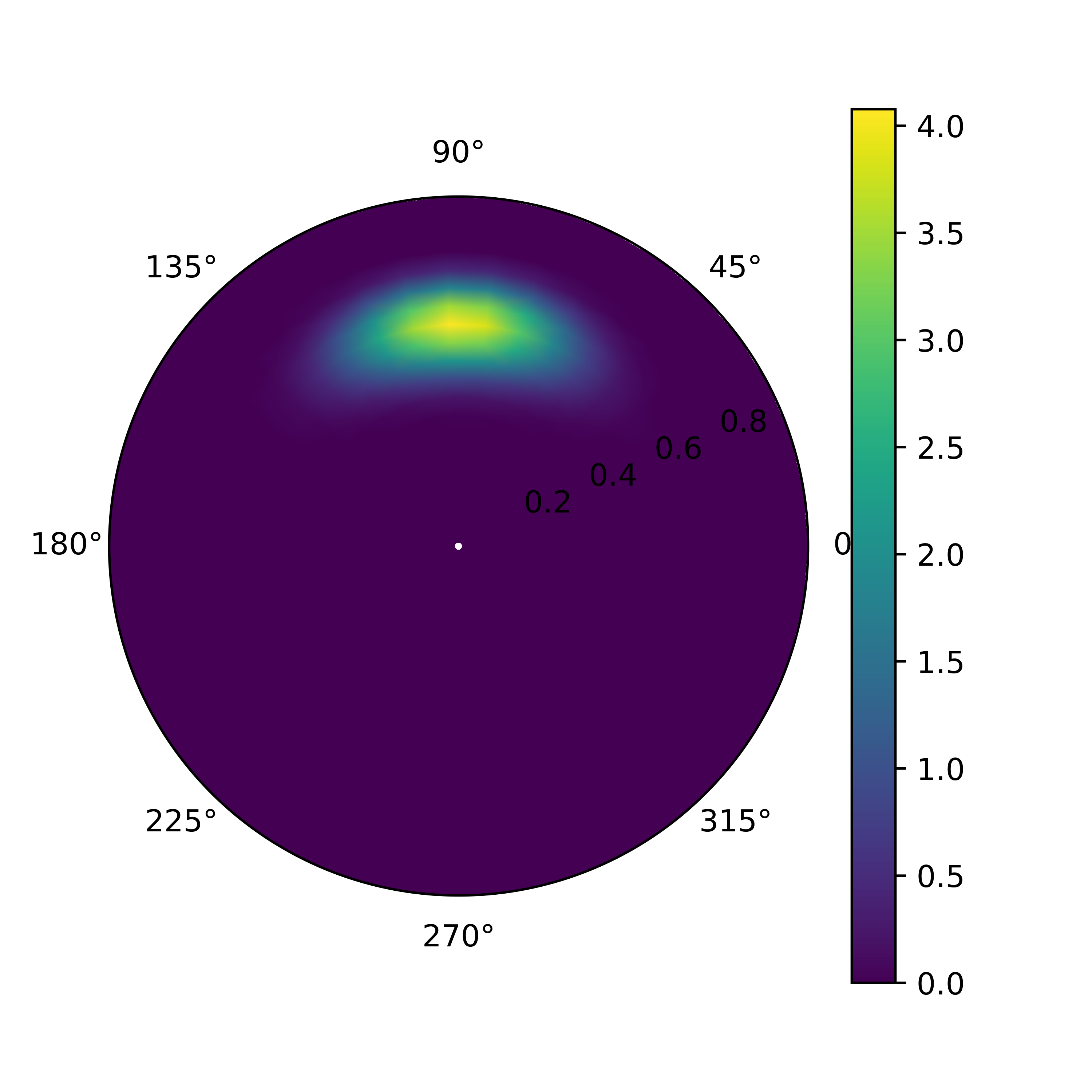}
       \caption{$P = 0.3\text{ c}$, $B_{12} = {1.5}$, $\chi = 89.3 \degree$.} 
    \end{subfigure}%

    \hspace{0.15cm}
        
    \begin{subfigure}[b]{0.3 \textwidth}
       \centering 
       \includegraphics[width=\textwidth]{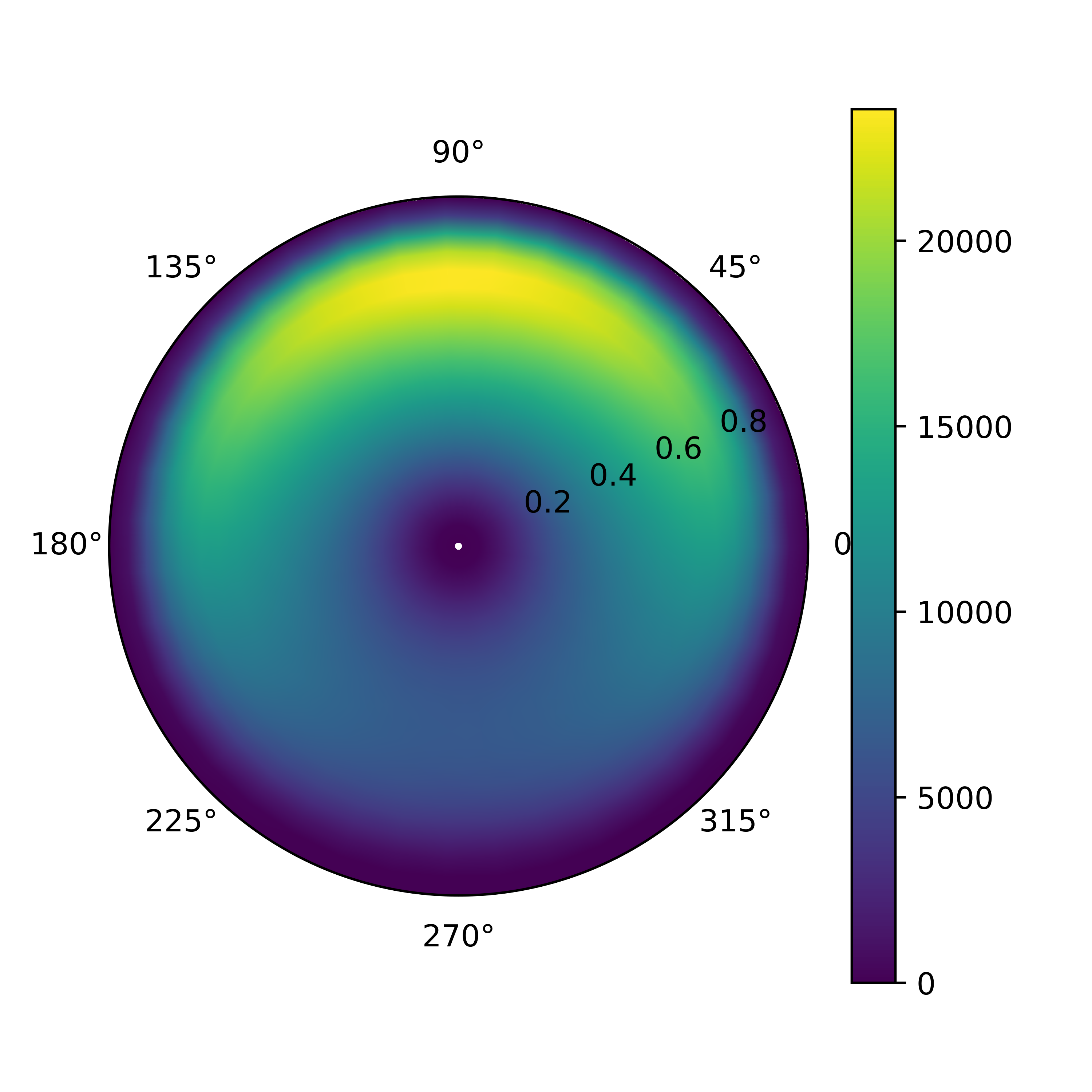}
       \caption{$P = 0.3\text{ c}$, $B_{12} = {7.0}$, $\chi = 85 \degree$.} 
    \end{subfigure}%
    \hspace{0.3cm}
    \begin{subfigure}[b]{0.3 \textwidth}
       \centering 
       \includegraphics[width=\textwidth]{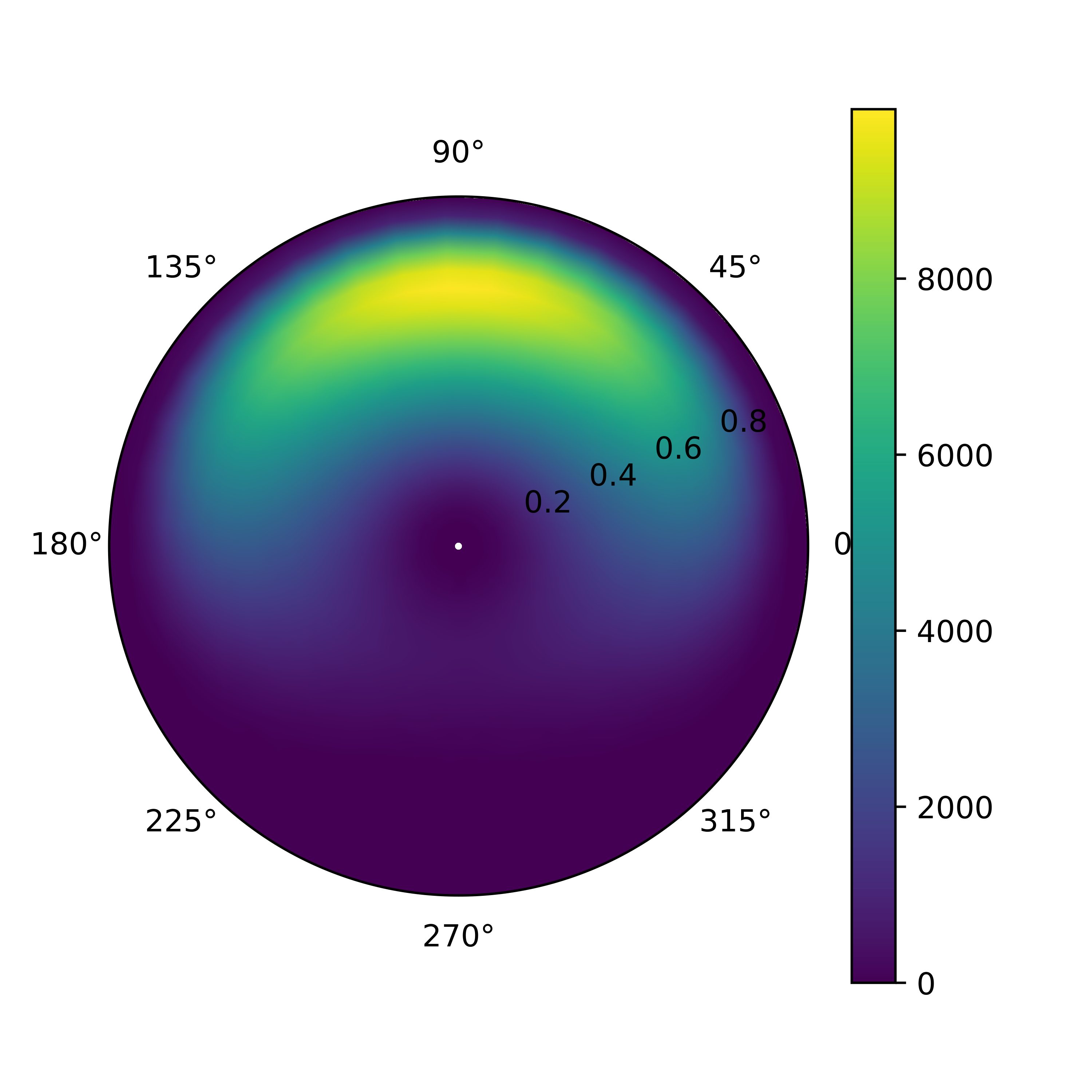}
       \caption{$P = 0.3\text{ c}$, $B_{12} = {7.0}$, $\chi = 88 \degree$.} 
    \end{subfigure}%
    \hspace{0.3cm}
    \begin{subfigure}[b]{0.3 \textwidth}
       \centering 
       \includegraphics[width=\textwidth]{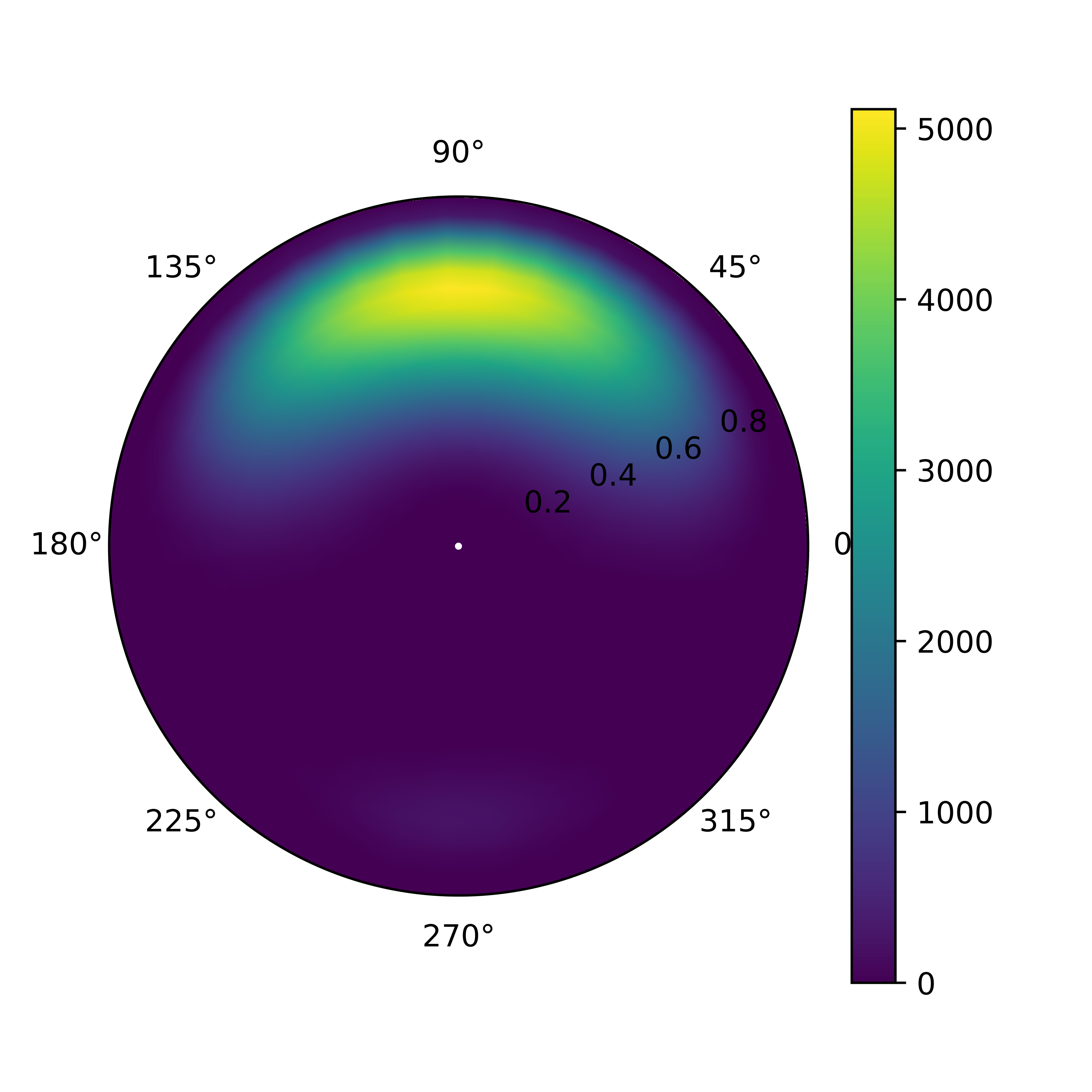}   
       \caption{$P = 0.3\text{ c}$, $B_{12} = {7.0}$, $\chi = 89.3 \degree$.} 
    \end{subfigure}%
   \caption{Plasma density distribution function $\lambda g(r_{\mathrm{m}}, \phi_{\mathrm{m}})$ for orthogonal pulsars for various magnetic fields and inclination angles.} 
   \label{fig:ns3D}
\end{figure*}

\subsection{Non-Axisymmetric case}

Fig.~\ref{fig:ns3D} shows plasma density distribution functions $\lambda g(r_{\mathrm{m}}, \phi_{\mathrm{m}})$ for various magnetic fields $B_{12}$ and inclination angles $\chi$ for period $P = 0.3$ s, typical for orthogonal interpulse pulsars. 
As can be seen in this figure, for moderate magnetic fields and angles that are not significantly different from 90\textdegree, plasma generation is greatly suppressed. At the same time, the density profiles generally follow the profiles described in earlier work~\cite{Novoselov}. Therefore, the main conclusion of that study regarding the statistics of interpulse pulsars remains valid: the observed number of orthogonal interpulse pulsars supports the law of the inclination angle $\chi$ evolution towards $\sim 90\degree$.

Recall that for orthogonal pulsars, the very possibility of generating secondary plasma is uncertain. Indeed, at  $\chi \sim 90 \degree$, the Goldreich-Julian density \eqref{eq:rho_GJ} is significantly lower than in the non-orthogonal case. As this value is the source term in the Poisson equation \eqref{eq:Poisson}, other things being equal, orthogonal pulsars should have a significantly lower accelerating potential, which could affect the possibility of plasma generation.
This question is particularly interesting in the context of the magnitude of the magnetic field required for secondary plasma generation, which is impossible to estimate without using an evolution model explicitly. Although for ordinary pulsars, choosing a specific model only leads to a difference of about a factor of $\sim 2-3$, for orthogonal pulsars, the difference may be significantly greater. While the generally accepted magnetohydrodynamics (MHD) model of evolution \citep{spitkovsky2006, kalcont2012, tchekhphil2016} predicts magnetic fields that are not significantly different from the standard estimate $B_{12} \approx \dot P_{-15}^{1/2} P^{1/2}$
\begin{equation}
    B_{12}^{\mathrm{MHD}}\approx \frac{\dot P_{-15}^{1/2} P^{1/2}}{\sqrt{1 + \sin^2{\chi}}},
    \label{eq:MHD}
\end{equation}
the BGI-model, proposed in the work~\citep{BGI93}, suggests that the magnetic field can be written as
\begin{equation}
    B_{12}^{\mathrm{BGI}}\approx \frac{\dot P_{-15}^{1/2} P^{1/2}}{\sqrt{\cos^2{\chi} + \mathcal{C}}},
    \label{eq:BGI}
\end{equation}
where $\mathcal{C} \sim \sqrt{\Omega R/c} \ll 1$. 
As a result, for orthogonal pulsars, the expression \ref{eq:BGI} leads to an estimate $B_{12} \approx 10 \, P^{3/4} \dot{P}_{-15}^{1/2}$ (see ~\cite{Novoselov}), which predicts magnetic field values several times higher than those predicted by the MHD model. At higher magnetic fields, the conditions for generating secondary particles in orthogonal pulsars are fulfilled over a wider range of periods, indicating the validity of the BGI model.

However, this issue requires a more in-depth analysis, which is beyond the scope of this paper. Accordingly, the purpose of this study was not to conduct calculations for specific radio pulsars. A separate article will focus on this topic. In particular, that is why we have limited ourselves to discussing the model of secondary plasma generation for magnetic fields that exceed standard estimates for interpulse pulsars but do not greatly exceed the limits of our model's applicability: $B < B_{\mathrm{cr}}$~\cite{IstSob}. We also propose to use this model to determine average profiles of observed radio emission in our next article.

\section{Conclusion}
\label{sec:Conclusion}
In the paper, a novel approach to determining acceleration potential was proposed. It is based on the concept of the vacuum gap, the height of which is assumed to be coordinate-dependent and is determined self-consistently with the accelerating potential. In order to implement this approach, an iterative procedure was constructed, where we alternately solve the Poisson equation in the gap and recalculate its height, until we converge to a certain solution. To do the former, Physics Informed Neural Network (PINN) method was successfully implemented. 

It should be emphasized that considered approach allowed us to quantitatively determine the non-axisymmetric structure of the accelerating potential for orthogonal pulsars for the first time. Using this method and the model of secondary electron-positron plasma generation, we calculated transverse secondary plasma density profiles both for ordinary and orthogonal pulsars. We also studied the effect of inverse Compton scattering and showed that even for the temperature of the polar cap of around  $T \sim 10^6 \, \text{K}$ this process can be neglected. 

Next, we have done preliminary study of the pair production efficiency for orthogonal pulsars. It was shown that for inclination angles $ 89 \degree \lesssim \chi \lesssim 91 \degree$ and moderate magnetic field $B_{12} \sim 2$, total pair multiplicity turns out to be several orders of magnitude smaller than that for the ordinary pulsars. At the same time, orthogonal pulsars are observed quite frequently \cite{Novoselov}, and estimates of their inclination angles fall  within the specified range.

Several explanations for this contradiction can be considered. First of all, as estimates of pulsars inclination angles  usually have big uncertainties, it is possible that we simply do not observe pulsars with such extreme inclination angles. Another explanation is that we underestimate their magnetic field magnitude. In this case, the evolution model should be refined. The possible non-dipolar structure of the magnetic field near the star, which can enhance secondary plasma generation, also cannot be ruled out. Therefore, this contradiction serves as a motivation to pay more attention to the luminosities of orthogonal pulsars and estimates of their inclination angles.

Finally, we want to note that one of the main applications of obtained results is the study of radiation propagation in neutron star magnetosphere. Indeed, as was shown in papers ~\citep{beskinphilippov2012, P&L2, Omode}, effects of refraction and absorption can significantly change the shapes of mean intensity profiles. Thus, it is impossible to verify radio emission models without knowledge about the real plasma density profiles. Moreover, the correct statistical treatment of orthogonal pulsars is impossible without the knowledge of their plasma density profiles, as the corresponding visibility function strongly depends on them~\citep{Novoselov}.

Obviously, it is important to keep in mind the model nature of the solved problem. Indeed, many numerical ~\citep{TimArons2013, TimHar2015, PhSC15} and analytical ~\citep{Tolman_Philippov_Timokhin} studies indicate a significant non-stationarity of secondary plasma generation. Thus, the limits of applicability of the proposed model require separate, detailed study. However, at present, there are no non-stationary models which are able to determine the spatial distribution of secondary plasma above the polar cap. So, it will be even more interesting to compare the results of this study with non-stationary numerical models, when their accuracy becomes sufficient to determine the density profiles of the outflowing plasma.

\section{Funding}
This work was supported by the Russian Science Foundation, project no. 24-22-00120.

\bibliographystyle{unsrt} 
\bibliography{references} 

\end{document}